%% file: Mink3S3_V2.tex
\documentclass[12pt]{article}

\pdfoutput=1

\usepackage[top=80pt,bottom=85pt,left=60pt,right=60pt]{geometry}
\usepackage{amssymb}
\usepackage{amsmath}
\usepackage{graphicx,color,subfigure}
\usepackage{float}
\usepackage{cite}
\usepackage{enumerate}
\usepackage{slashed}
\usepackage{bbm}
\usepackage[debug,pageanchor=false]{hyperref}
\definecolor{link}{rgb}{.8,.15,.1}
\hypersetup{colorlinks=true,linkcolor=link,citecolor=link,urlcolor=link,linktocpage}

\makeatletter
\@addtoreset{equation}{section}
\makeatother

\input{niall-macros.tex}

\input{daniel-macros.tex}

\begin{document}


\begin{titlepage}

\begin{center}

\vskip .5in 
\noindent

{\Large \bf{Mink$_3\times S^3$ solutions of type II supergravity}}

\bigskip\medskip

Niall T. Macpherson$^{a,b,c}$, Jes\'{u}s Montero$^{a,d}$, Dani\"{e}l Prins$^{a,e}$

\bigskip\medskip
{\sl $^a$ Dipartimento di Fisica Universit\`{a} di Milano-Bicocca, \\
I-20126 Milano, Italy}
\vskip 0.1cm
{\sl $^b$ INFN, sezione di Milano-Bicocca,
 I-20126 Milano, Italy}
 \vskip 0.1cm
{\sl $^c$ SISSA International School for Advanced Studies and INFN, sezione di Trieste,\\ 34136, Trieste, Italy}
\vskip 0.1cm
{\sl $^d$ Department of Physics, University of Oviedo, Avda. Calvo Sotelo 18, 33007 Oviedo, Spain}
\vskip 0.1cm
 {\sl $^e$ Institut de physique th\'eorique, Universit\'e Paris Saclay, CNRS, CEA \\F-91191 Gif-sur-Yvette,
France}
\vskip 0.1cm

\vskip 0.2 cm
\vskip 0.2 cm
{\textsf{nmacpher@sissa.it}} \\
{\textsf{monteroaragon@uniovi.es}}\\
{\textsf{daniel.prins@cea.fr}}
\vskip .1in
\end{center}

\noindent
\begin{center}
{\bf ABSTRACT }
\end{center}
\noindent We initiate the classification of supersymmetric solutions of type II supergravity on $\mathbbm{R}^{1,2} \times S^3 \times M_4$. We find explicit local expressions for all backgrounds with either a single Killing spinor or two of equal norm, up to PDE's. We show that the only type II AdS$_4\times S^3$ solution is the known $\mathcal{N}=4$ AdS$_4$ background obtained from the near-horizon limit of  intersecting D2-D6 branes. Various known branes and intersecting brane systems are recovered, and we obtain a novel class of $\mathbbm{R}^{1,2} \times S^2\times S^3$ solutions in IIA.
\\\\

\vfill
\eject

\end{titlepage}


\tableofcontents

\section{Introduction}

The advent of the AdS-CFT correspondence has led to significant interest in the construction of Anti-de Sitter string backgrounds in various dimensions and with various amounts of supersymmetry. One of the most famous AdS$_4$ backgrounds is the $\text{AdS}_4 \times \mathbb{CP}^3$ solution. The discovery of this solution far pre-dates the correspondence \cite{np}, however it was not realised how it fit into the holographic paradigm until the works of \cite{bl} and \cite{ABJM}. A plethora of other such AdS$_4$ classes and explicit examples have been found using (in some cases vastly) different methods and exhibiting different amounts of supersymmetry: consider the very incomplete list of \cite{lt1, blt, klt, klpt,tom2, ckkltz, lt2, bc, varela,Rota:2015aoa, grig} for $\mathcal{N} =1$, \cite{lt3, pz, ym, gjv, gpt, gmps, ptz} for $\mathcal{N} = 2$, \cite{pr} for $\mathcal{N} = 3$ and \cite{deg1, deg2,abeg1, abeg2} for $\mathcal{N} = 4$. Solutions with $\mathcal{N}> 4$  where recently classified in \cite{Haupt:2017bnj}, they are very restricted.

One of the more prominent methods of finding  AdS$_d$ backgrounds is to find bosonic solutions with an AdS$_d$ factor to the supersymmetry constraints, which also satisfy the Bianchi identities. As a consequence of various integrability theorems, such solutions automatically solve the equations of motions. The Killing spinor equations reduce to constraints on the internal manifold, which can then be solved by means of $G$-structure and generalised geometrical techniques. The literature usually approaches this problem by assuming an AdS$_d$ from the start.  However we are also interested in solutions of relevance to flux compactifications and the broader definition of holography that includes non conformal solutions. As such we shall consider assume Minkowski factor, in this case  Mink$_3$, so that our results are more broadly applicable.\\

Finding Minkowski solutions using $G$-structure techniques \cite{gmw, gkmw,gmpt,hlmt} or otherwise is by now quite a mature program, see \cite{and, abr, his, Andriot:2017jhf} for some recent examples. Usually the aim is to preserve minimal or even no supersymmetry for phenomenological reason which makes the problem in general quite hard. We shall  take inspiration from \cite{Macpherson:2016xwk} and assume the existence of an $S^3$ factor in the metric. This will necessarily mean that we are dealing with at least $\mathcal{N}=2$ which is of less phenomenological interest, however with these solutions classified it should then be possible to systematically break some (or even all) of this symmetry by deforming the $S^3$.\\

In this paper we classify all supersymmetric solutions of Type II supergravity on $\rbb^{1,2} \times S^3 \times M_4$, under the assumption that the seven-dimensional internal Killing spinors have equal norms and that the physical fields of the solution respect the $ISO(1,2) \times SO(4)$ isometry subgroup. Our classification is quite detailed, going as far as to give explicit local expressions the metric, fluxes and dilaton in terms of simple (Laplace-like) PDE's. As we shall see, solutions in this class are generically $\mathcal{N}=2$, from the Minkowski perspective, and support a $SU(2)$ R-symmetry realised geometrically as one factor of the $SO(4) \simeq SU(2)_+ \times SU(2)_-$ isometry group manifold of $S^3$ - the remaining $SU(2)$ factor is a  "flavour" under which the Killing spinors are uncharged.\footnote{As such all classes we present are compatible with performing non-Abelian T-duality \cite{oq,st,inst} on this $SU(2)$ whilst preserving $SU(2)_R$ \cite{kymc}}
This may sound strange as there is no 3d superconformal algebra with $SU(2)_R$, but this only matters for solutions where $\rbb^{1,2}$ is part of a $AdS_4$ factor so that $SO(2,3)$ is realised. Ultimately our results end up side stepping this issue as in general their is either an enhancement of the R-symmetry to $SO(4)$ via the emergence of an additional $S^{2,3}$ factor, or an enhancement of the Minkowski factor to dimensions where $SU(2)_R$ is a necessary part of the superconformal algebra.\\

The classification recovers various well-known intersecting brane systems listed in \cite{youm} and some of their U-duals and some of their S-duals. New classes we find include a pure NS $\rbb^{1,2} \times S^3 \times S^2 \times \rbb$ vacuum, its U-dual in IIB, the cone over $\rbb^{1,2} \times S^3 \times S^3$, and a novel class of $\mathbb{R}^{1,2}\times S^2\times S^3\times \Sigma_2$ solutions in massless and massive IIA.\\

One of our main results is that the only compact $\text{AdS}_4 \times S^3 \times M_3$ solution of type II supergravity is the known  $\ncal=4$ solution of type IIA on a foliation of $\text{AdS}_4 \times S^3 \times S^2$ over an interval which is the near-horizon limit of the an D2-D6 brane system. The required  $SO(4)_R$ is realised with one $SU(2)$ from each sphere, and not the $S^3$ alone. Indeed this is to be expected as if the 3-sphere does realise two $SU(2)$ R-symmetries there would be two sets of $\mathcal{N}=2$ spinors  transforming in the $(2,1)$ and $(1,2)$ of $SO(4)$ -  there is no $\mathcal{N}=4$  super-conformal algebra in 3d with $Q$-generators that transform in this fashion. So it seems likely that the only avenue left open for holographic duals of $\mathcal{N}=4$ is to seek AdS$_4\times S^2\times S^2$ solution like \cite{deg1, deg2}, but in massive IIA.\\

Our other main result is the discovery of a new class of $\mathcal{N}=4$ solutions on $\mathbb{R}^{1,2}\times S^2\times S^3\times \Sigma_2$ preserving an $SO(4)$ R-symmetry but no AdS$_4$ . These generically have all possible IIA fluxes turned on and can be divided into cases either in massless or massive IIA at which point solutions are in one to one correspondence with a single PDE on $\Sigma_2$.  In particular the massless solutions are governed by a 3d cylindrical Laplace equation with axial symmetry. These classes look very promising both for finding compact Mink$_3$ solutions, but also possibly solutions that asymptote to AdS.\\

Let us now describe the outline of the paper:
in order to solve the supersymmetry constraints, we will make use of the reformulation of the Killing spinor equations in terms of so-called pure spinor equations. Such pure spinor equations were first used for backgrounds of the form $M_{10} = \rbb^{1,3} \times M_6$, where it was shown that they are related to integrability constraints of generalised almost complex structures on the internal space $M_6$ \cite{gmpt}. For backgrounds of the form $M_{10} = \rbb^{1,2} \times M_7$, the pure spinor equations were constructed in \cite{hlmt} (see also \cite{tom}). Next, we decompose $M_7 = S^3 \times M_4$, leading to pure spinor equations on the internal $M_4$. We explain this setup in detail in section \ref{setup}. The resulting supersymmetry constraints vary significantly, depending on whether the theory at hand is type IIA or type IIB. We will solve the supersymmetry constraints as well as the Bianchi identities for IIB backgrounds in section \ref{IIB} and for IIA backgrounds in section \ref{IIA}. In section \ref{sec:ads4}, we then show that there is a unique solution with a warped AdS$_4$ factor, obtained from the D2-D6 system . In addition to the case where the internal Killing spinors have equivalent norm, in section \ref{e20} we examine all backgrounds in the case where one of the Killing spinors vanishes, i.e., $\e_2 = 0$. In this case, there is no need to distinguish between IIA and IIB; we demonstrate that all such backgrounds are pure NSNS and give the solutions.
In the appendix, we discuss conventions and identities used, a mild extension of the 3+7 pure spinor equation construction (including the non-equivalent norm case), and a discussion on similar backgrounds from an M-theory perspective.

\section{Mink$_3$ with an $S^3$ factor}\label{setup}
We are interested in solutions to type II with at least a three-dimensional external Minkowski component, with the fluxes respecting  the three-dimensional Poincar\'{e} invariance:
\beq\label{eq:main_Ansatz}
ds^2= e^{2A}ds^2(\mathbb{R}^{1,2})+ ds^2(M_7)~,~~~
F= f+ e^{3A}\textrm{Vol}_3\wedge\star_7 \lambda(f)~,
\eeq
where the RR flux $f$ is a polyform on $M_7$ and the warp factor $A$ and the dilaton $\Phi$ are functions on $M_7$.\footnote{We work in the democratic formalism. Other conventions can be found in appendix \ref{conv}} Moreover, we take the NSNS 3-form $H$ to be internal as well.
The Killing spinors for $\mathcal{N} = 1$ supersymmetric solutions decompose as
\eq{\label{10dks}
\e_1 = \left(\begin{array}{c} 1 \\   - i \end{array} \right) \otimes \zeta \otimes \chi_1 ~,~~~~
\e_2 = \left(\begin{array}{c} 1 \\ \pm i \end{array} \right) \otimes \zeta \otimes \chi_2~,
}
where $\zeta$ is a Majorana spinor of $Spin(1,2)$ and $\chi_{1,2}$ are Majorana spinors of $Spin(7)$ and where the upper (lower) signs are taken in IIA (IIB). Following \cite{hlmt}, we define two real seven-dimensional bispinors $\Phi_{\pm}$ in terms of $\chi_{1,2}$:
\beq\label{bispinor}
\Phi_+ + i \Phi_-= 8e^{-A}\chi_{1}\otimes \chi_2^{\dag}~,
\eeq
where the subscript $+/-$ refers to the even/odd forms in the decomposition of the polyform. The conditions for unbroken $\mathcal{N}=1$ supersymmetry are equivalent to
\begin{subequations}\label{7dsusy}
\begin{align}
&d_H(e^{2A-\Phi}\Phi_{\pm})=0~,\\[2mm]
&d_H(e^{3A-\Phi}\Phi_{\mp})+  e^{3A} \star_7 \lambda(f)=0~,\\[2mm]
&(\Phi_{\pm}\wedge \lambda(f))\big\lvert_{\text{Top}}=0~,
\end{align}
\end{subequations}
as long as the norms of spinors $\chi_{1,2}$ are equal\footnote{In \cite{hlmt}, \cite{tom} an additional constraint that was imposed in order to derive \eqref{7dsusy} was that the external component of the NSNS 3-form flux is trivial; unlike in four dimensions, this is not enforced by Poincar\'{e} invariance. It turns out that this second assumption is redundant though, as is shown in appendix \ref{susyconditions}: if $|\chi_1|^2 = |\chi_2|^2$ and spacetime does not admit a cosmological constant, then supersymmetry enforces that the external NSNS flux vanishes.}, which leads to
\beq
|\chi^1|^2=|\chi^1|^2=e^{A}
\eeq
The assumption of equal norm is a global requirement for AdS$_4$ (see footnote 6 in section \ref{sec:ads4}), and a local requirement for the existence of calibrated D-branes or O-planes (see section \ref{e20}), however this is not a requirement in general - rather we view this as a well-motivated simplifying assumption.\\

Next, we require that the internal space can be decomposed locally as $M_7 = S^3 \times M_4$, and in order to ensure that compactification leads to an $SO(4)$ global symmetry we insist that the fluxes respect the $SO(4)$ isometry. As a result, the metric and fluxes decompose further as
\beq\label{734}
ds^2(M_7)= e^{2C}ds^2(S^3) + ds^2(M_4)~,~~~
 f= G_{\mp}+ e^{3C}\text{Vol}(S^3)\wedge G_{\pm}~,~~~
 H= H_3+ H_0e^{3C}\text{Vol}(S^3)~.
\eeq
We decompose the 7d spinors in the same fashion in terms of a single\footnote{As explained in Appendix \ref{sec: su2doublet}, there are two independent types of Killing spinors on $S^3$, $\xi_+$ and $\xi_-$ - however they cannot be mapped to each other using the $SO(4)$ invariants of the fluxes or the Killing spinor equations. This is all that appears when one decompose $M_7=S^3\times M_4$, so if one were to include terms like $\xi_+\otimes \eta_+$ and $\xi_-\otimes \eta_-$ then  reduced the 7d spinor conditions to 4d ones you would find that $\eta_{\pm}$ never mix. So setting one of $\eta_{\pm}$ to zero excludes no solutions in our analysis.}  pseudoreal (i.e., $(\xi^c)^c = - \xi$) Killing spinor $\xi$ on $S^3$, and two pseudoreal spinors $\eta_{1,2}$ on $M_4$:
\beq\label{eq:general7sspinrs}
\chi_i= e^{\frac{A}{2}}(\xi\otimes \eta_i+  \xi^c\otimes \eta^{c}_i) =e^{\frac{A}{2}}\xi^a\otimes \eta^a_i~,\qquad \qquad i = 1,2
\eeq
which is the most general parameterisation consistant with an $S^3\times M_4$ product and the  Majorana condtion.\footnote{One might imagine it was possible to construct a more general 7d spinor from two 4d spinors like $\xi\otimes \eta+  \xi^c\otimes \tilde{\eta}$. But if one then adds the Majorana conjugate to this the resulting spinor can be put  in the form of \eqref{eq:general7sspinrs} by redefining $\eta,~\tilde{\eta}$.} Note that we do not restrict the $Spin(4)$ spinors $\eta_i$ to be chiral and we normalise $\eta^{\dag}_{1,2}\eta_{1,2}=1$~. The Killing spinors on $S^3$ satisfy the Killing spinor equation
\eq{\label{s3ks}
\nabla_\a \xi = \frac12 i \nu \s_\a \xi ~, ~~~~ \nu = \pm 1~.
}
which preserves two supercharges for each of $\nu=\pm 1$. We will not make a choice of $\nu$ so we can establish whether any solutions are independent of this choice - the $S^3$ of such a solution would preserve 4 supercharges. As explained in Appendix \ref{sec: su2doublet} a spinor on $S^3$ defines a doublet
\beq
\xi^a=\left(\begin{array}{c}\xi\\\xi^c\end{array}\right)
\eeq
which is charged under one $SU(2)$ factor of $SO(4)=SU(2)_+\times SU(2)_-$, depending on the sign of $\nu$ - $\xi^a$ is a singlet under the action of the second $SU(2)$. As such, a generic solution with Mink$_3\times S^3$ will have an R-symmetry $SU(2)_R$ and an additional global flavour symmetry $SU(2)_F$. Such solutions preserve at least  $\mathcal{N}=2$ supersymmetry from the 3d perspective, so 4 real supercharges - indeed, the 10d Killing spinors may be written as
\eq{\label{10dks}
\e_1 = \left(\begin{array}{c} 1 \\   - i \end{array} \right) \otimes \zeta^{a} \otimes (\xi^{a}\otimes \eta^1+\xi^{a c}\otimes \eta^{1c}) ~,~~~~
\e_2 = \left(\begin{array}{c} 1 \\ \pm i \end{array} \right) \otimes \zeta^{a} \otimes (\xi^{a}\otimes \eta^2+\xi^{a c}\otimes \eta^{2c})~,\nn
}
where $\zeta^{a}$ is a doublet of Killing spinors on $\rbb^{1,2}$, that allow the 10d spinors to be invariant under $SU(2)_R$ transformations. However we only need to solve an $\mathcal{N}=1$ sub-sector, because the part of the Killing spinor which couples to $\zeta^1$ is mapped to the part coupling to $\zeta^2$ under the action of $SU(2)_R$ - so if you solve one part, the other is guaranteed. If a solution ends up being independent of $\nu$ then there is a copy of \eqref{10dks} for each sign and supersymmetry is doubled to $\mathcal{N}=4$ - there are two $SU(2)$ R-symmetries, but they do not appear as a product so do not form $SO(4)_R$ - as we shall see, this only happen in a small number of special cases.\\

Using the gamma matrix decomposition \eqref{gammadecomp}, the seven-dimensional bispinor \eqref{bispinor} decomposes as
\beq\label{7dbispinordecomp}
\chi_1\otimes \chi_2^{\dag}= (\xi^a\otimes \xi^b)_+\wedge (\eta^a\otimes\eta^b)+(\xi^a\otimes \xi^b)_+\wedge (\hat\gamma\eta^a\otimes\eta^b).
\eeq
Here, $\hat{\gamma}$ is the four-dimensional chirality matrix and the $\pm$ subscripts again refer to even and odd form components. We see that the components are in fact matrices and that the seven-dimensional bispinor is constructed as the trace of the product of the components.\\

The $S^3$ component leads to the bispinor matrix
\beq\label{s3bispinor}
\xi^a\otimes \xi^{b\dag}=\frac{1}{2}\bigg((1- i  e^{3C}\text{Vol}(S^3))+ \frac{1}{2}\big(e^{C}K_i- \frac{\nu}{2} i e^{2C}dK_i\big)(\sigma^i)_{ab}\bigg) \;,
\eeq
where $K_i$ is a vielbein defining a trivial structure on $S^3$ (see appendix \ref{conv}).\\

The $M_4$ component leads to the bispinor matrix
\beq\label{m4bispinor}
(\eta^{a}_1\otimes\eta^{b\dag}_2)_{\pm}=\left(\begin{array}{cc} \psi^1_{\pm}&\psi^2_{\pm}\\
\mp(\psi^2_{\pm})* &\pm(\psi^1_{\pm})* \end{array}\right),~~~~(\hat\gamma\eta^{a}_1\otimes\eta^{b\dag}_2)_{\pm}=\left(\begin{array}{cc} \psi^1_{\hat\gamma\pm}&\psi^2_{\hat\gamma\pm}\\
\mp(\psi^2_{\hat\gamma\pm})* &\pm(\psi^1_{\hat\gamma\pm})* \end{array}\right),
\eeq
where
\beq
\psi^1= 4\eta^1\otimes \eta^{2\dag},~~~\psi^2= 4\eta^1\otimes \eta^{2c\dag},~~~\psi^1_{\hat\gamma}= 4\hat\gamma\eta^1\otimes \eta^{2\dag},~~~\psi^2_{\hat\gamma}= 4\hat\gamma\eta^1\otimes \eta^{2c\dag}
\eeq
Since the matrix entries are somewhat involved, we refer to appendix \ref{4dbispinors} for details.
Plugging both components \eqref{s3bispinor}, \eqref{m4bispinor} into the seven-dimensional bispinors \eqref{7dbispinordecomp}, it follows that
\begin{align}\label{eq:7d-bispinors}
\Phi_+&= \text{Re}\psi^1_+ - e^{3C}\text{Vol}(S^3)\wedge\text{Im}\psi^1_{\hat\gamma-}+ \frac{e^{C}}{2}(K_1\wedge \text{Re}\psi^2_{\hat\gamma_-}+ K_2\wedge \text{Im}\psi^2_{\hat\gamma-}+ K_3\wedge\text{Re}\psi^1_{\hat\gamma_-} ),\nn\\[2mm]
&-\frac{e^{2C}}{4}(K_1\wedge K_2\wedge \text{Im}\psi^1_++K_1\wedge K_3\wedge \text{Re}\psi^2_++K_2\wedge K_3\wedge \text{Im}\psi^2_+),\\[2mm]
\Phi_-&=\text{Im}\psi^1_- - e^{3C}\text{Vol}(S^3)\wedge\text{Re}\psi^1_{\hat\gamma+}+ \frac{e^{C}}{2}(K_1\wedge \text{Im}\psi^2_{\hat\gamma_+}- K_2\wedge \text{Re}\psi^2_{\hat\gamma+}+ K_3\wedge\text{Im}\psi^1_{\hat\gamma_+} ),\nn\\[2mm]
&+\frac{e^{2C}}{4}(K_1\wedge K_2\wedge \text{Re}\psi^1_--K_1\wedge K_3\wedge \text{Im}\psi^2_-+K_2\wedge K_3\wedge \text{Re}\psi^2_-).\nn
\end{align}
At this point, the IIA and IIB supersymmetry equations diverge, and we shall relegate their explicit form to the relevant sections.\\

With our set up, a solution to the supersymmetry equations is a solution to the equation of motion if and only it satisfies the Bianchi identites \cite{lt1} \cite{ggpr} \cite{Gauntlett:2004zh}. These  are given by  $d_H F = d H = 0$ away from localised sources. By definition, a localised (magnetic) source manifests itself in the Bianchi identity of some field strength $F$ as $d F = Q \delta^n (x)$ and hence in such cases $F$ is discontinuous. Loosely speaking, a localised source corresponds physically to an extended object (such as a brane) located at a submanifold of the ten-dimensional spacetime $\mathcal{S} \subset M_{10}$ which is pointlike in some of the local coordinates. The standard approach to obtaining backgrounds, which we follow as well, is to first solve the supersymmetry equations by introducing local coordinates, and then afterwards determine the physically sensible range of these local coordinates by examining the obtained geometry and fluxes.
The presence of localised sources is signified by discontinuities of not just the fluxes, but of the spacetime geometry as well, precisely at the location of the sources. Therefore, it is possible to obtain solutions with localised sources even when making use of the Bianchi identities with no sources: one examines possible discontinuities in the geometry and fluxes and determines whether or not such discontinuities are associated with localised sources or not by comparing them with the divergent behaviour of known extended objects.

Making use of the flux decomposition \eqref{eq:main_Ansatz}, \eqref{734}, the Bianchi identies thus reduce to
\eq{
d_{H_3} \big( e^{3A + 3C} \star_4 \l(G_\pm) \big) =
d_{H_3} \big( e^{3C} \star_4 \l(G_\mp) \big) &= 0 ~, \\
d_{H_3} (G_\pm) =
d_{H_3} \big( e^{3C} G_\mp \big) &= 0 ~,\\
d H_3 &= 0 ~.
}
This is after imposing $H_0=0$, which turns out to be a requirement for every solution to the supersymmetry equations that we obtain.

\subsection{Summary of obtained backgrounds}\label{summary}
As the rest of the paper is somewhat technical, let us summarise our results here. We find a number of well-known backgrounds, as well as some new ones. \\

In type IIB with internal Killing spinors of equal norm, we find:
\begin{enumerate}
\item The intersecting D3-D7 system with metric \eqref{d3d71}, fluxes \eqref{d3d72} and scalar field constraints \eqref{d3d73}.
\item The D5-brane with metric \eqref{d51}, fluxes \eqref{d52} and scalar field constraints \eqref{d53}.
\item A generalization of the D5-brane generated by U-duality. The metric is given by \eqref{d5u1}, the fluxes by \eqref{d5u2}, scalar field constraints by \eqref{d5u3}.
\item A new background on the cone over $\rbb^{1,2} \times S^3 \times S^3_{\text{sq}}$, with $S^3_{\text{sq}}$ a generically squashed three-sphere admitting an $SU(2) \times U(1)$ isometry group. For the unsquashed limit, the metric is given by \eqref{eq: d5thing}, the fluxes by \eqref{d5s2}, the scalar field constraints by \eqref{d5s3}.
In the generic squashed case, the metric and dilaton are given by \eqref{eq:newIIB1}, the fluxes by \eqref{eq:newIIB2}. We note that the more general squashed case can be obtained from the unsquashed case by a duality chain.
\end{enumerate}

In type IIA with internal Killing spinors of equal norm, we find:
\begin{enumerate}
\item The intersecting D4-D8 system with metric \eqref{d4d81}, fluxes \eqref{eq:branch1IIAbeta0Fluxes} and scalars constraints \eqref{d4d83}, \eqref{eq:branch1IIAbeta0PDEs}.
\item The intersecting D2-D6 system with metric \eqref{d2d61}, fluxes \eqref{d2d62} and scalar constraints \eqref{d2d64}, \eqref{d2d63}.
\item A generalization of the D4-D8 system generated by U-duality. The metric is given by \eqref{d4u1}, the fluxes by  \eqref{d4u2b}, and scalar constraints by \eqref{d4u3}.
\item A class of new backgrounds. The metric contains an $\rbb^{1,2} \times S^3 \times S^2$ factor, with various warpings, and is given by \eqref{nw1}. The warp factors are constrained by various PDE, given in \eqref{nw3}. In general, all fluxes are turned on and are given by \eqref{eq:lastfluxes}. This new class of backgrounds contains a subset with a $U(1)$ isometry. In this case, T-dualising along the isometry direction leads to the new IIB backgrounds outlined above, with generic squashing.
\end{enumerate}

In addition, we find two more backgrounds when setting $\e_2 = 0$. These backgrounds are pure NS, and as such, can be found in both type IIA and type IIB. We find:
\begin{enumerate}
\item The NS5-brane, with metric \eqref{ns51}, flux \eqref{ns52} and the scalar constraints \eqref{ns53}.
\item A pure NS background on $\rbb^{1,2} \times \rbb \times S^3 \times S^3$, dual to the new (unsquashed) IIB background. The metric is given by \eqref{e20s3s310}, the flux by \eqref{e20s3s32}. All scalars are determined up to constant factors.
\end{enumerate}

\section{Mink$_3$ with an $S^3$ factor in IIB}\label{IIB}
The type IIB supersymmetry equations are obtained by plugging the decomposed seven-dimensional bispinors \eqref{eq:7d-bispinors} into the seven-dimensional supersymmetry constraints \eqref{7dsusy}. This leads to the following constraints on the four-dimensional bispinors
\begin{subequations}\label{eq: IIbSUSY}
\begin{align}
&d_{H_3}(e^{2A-\Phi}\text{Re}\psi^1_+)=0~,\label{eq: IIBSUSYa}\\[2mm]
&d_{H_3}(e^{3A+2C-\Phi}\psi^2_-)+ 2i\nu e^{3A+C-\Phi}\psi^2_{\hat\gamma+}=0~,\label{eq: IIBSUSYb}\\[2mm]
&d_{H_3}(e^{2A+2C-\Phi}\psi^2_+)+ 2i\nu e^{2A+C-\Phi}\psi^2_{\hat\gamma-}=0~,\label{eq: IIBSUSYc}\\[2mm]
&d_{H_3}(e^{3A+2C-\Phi}\text{Re}\psi^1_-)- 2\nu e^{3A+ C-\Phi}\text{Im}\psi^1_{\hat\gamma+}=0~,\label{eq: IIBSUSYd}\\[2mm]
&d_{H_3}(e^{2A+2C-\Phi}\text{Im}\psi^1_+)+ 2\nu e^{2A+ C-\Phi}\text{Re}\psi^1_{\hat\gamma-}=0~,\label{eq: IIBSUSYe}\\[2mm]
&d_{H_3}(e^{2A+ 3C-\Phi}\text{Im}\psi^1_{\hat\gamma-})+ e^{2A+ 3C-\Phi}H_0 \text{Re}\psi^1_+=0\label{eq: IIBSUSYf},
\end{align}
\end{subequations}
while the fluxes are determined by
\begin{subequations}
\begin{align}
&d_{H_3}(e^{3A-\Phi}\text{Im}\psi^1_-)+ e^{3A}\star_4 \lambda(G_+)=0~,\\[2mm]
&d_{H_3}(e^{3A+3C-\Phi}\text{Re}\psi^1_{\hat\gamma_+})-e^{3A+3C-\Phi}H_0 \text{Im}\psi^1_-+ \nu e^{3A+3C}\star_4 \lambda(G_-)=0
\end{align}
\end{subequations}
and must additionally satisfy the pairing equation
\beq
\bigg(\text{Im}\psi^1_{\hat\gamma-}\wedge\lambda(G_-)-\text{Re}\psi^1_+\wedge \lambda(G_+)\bigg)\bigg \lvert_{4}=0~.
\eeq
In order to solve these, we will first examine the 0-form conditions, These are given by
\eq{
(\psi^2_{\hat\gamma})_0 = (\text{Im}\psi^1_{\hat\gamma})_0= H_0(\text{Re}\psi^1_{\hat\gamma})_0 = 0 \;,
}
We solve the first two of these in Appendix \ref{4dbispinors},  which leads to a spinor ansatz depending on 6 real functions with support on $M_4$
\beq
\alpha,~~a_1,~~~b_1,~~~\l_1,~~~\l_2~~~,\l_3
\eeq
subject to the constraint
\beq
a_1^2+b_1^2+\l_1^2+\l_2^2+\l_3^2=1.
\eeq
The third 0-form constraint, which is unique to IIB, still needs to dealt with. After making use of \eqref{eq:4dbispinors}, it reduces to
\beq\label{eq:zeroformsIIB}
H_0(a_1 \cos\frac{\alpha}{2}+ b_1 \sin\frac{\alpha}{2})=0~.
\eeq
 Here, as well as in IIA, the solutions depend drastically on the behaviour of $\alpha$. We can distinguish between three different cases: $\a = 0$, $\a = \frac12 \pi$, and generic $\a \in (0, \pi), \a \neq \frac12 \pi$. Let us reiterate that we introduced $\a$ in \eqref{eq: general4dspinors} by defining
\eq{
\eta_1 = \cos \left(\frac{\a}{2} \right) \eta + \sin \left( \frac{\a}{2} \right) \hat{\gamma} \eta~,
}
where $\eta$ is a locally defined non-chiral spinor, where the chiral components are normalised. Note that the non-chirality is crucial: it ensures that $\eta$ can be used to define the local trivial structure (i.e., the vielbein) via \eqref{eq: canonicalframe}. In the case that $\a = 0$, the 4d internal Killing spinors $\eta_1 = \eta$ are such that the chiral components of $\eta_1$ have equal norm. In the case that $\a = \pi / 2$, we see that $\eta_1$ becomes chiral. It turns out that we can treat this case together with $\a \neq 0$, but find no such solutions. Thus we seperate our solutions into two branches.\\

\textbf{Branch I:} Here $\alpha=0$. The only non-trivial zero form is $a_1 H_0=0$, which a priori can be solved in two ways. However, we shall see in the next section that only $H_0=0$ is consistent with the higher form conditions.
In order to solve \eqref{eq:sum of squares} we parametrise
\beq\label{eq: branchoneparameterisation}
a_1=\sin\beta,~~b_1= \cos\beta \sin\delta,~~ \l_1= y_1\cos\beta\cos\delta ,~~ \l_1= y_2\cos\beta\cos\delta ,~~ \l_3= y_3\cos\beta\cos\delta ,
\eeq
with
\beq\label{eq:S2embedding}
y_1=\sin\theta\cos\phi,~~~y_2=\sin\theta\sin\phi,~~~y_3=\cos\theta.
\eeq


\textbf{Branch II:} Here $0<\alpha<\pi$. Note that $\a = \pi$ is equivalent to $\alpha=0$, which is easiest to see by sending $\eta\to \hat\gamma \eta$ in \eqref{eq: general4dspinors}. We choose to parametrise
\begin{subequations}\label{eq:branchIIparamaterisation}
$$\begin{alignedat}{6}
a_1&=\cos\beta \sin(\delta-\frac{\alpha}{2}) &~,~~~~ & b_1~ &=& \cos\beta \cos(\delta-\frac{\alpha}{2})\\
\l_1&=-\cos\beta \cos\delta y_1 &~,~~~~ & \l_2~ &=&  - \sin\beta y_3 \\
\l_3&=-\sin\beta y_2      &~~~~~ & ~&& ~\\
\end{alignedat}
$$\end{subequations}
where $y_i$ is defined in \eqref{eq:S2embedding}. This ensures \eqref{eq:sum of squares} and all but the first equation of \eqref{eq:zeroformsIIB} are solved, which becomes
\eq{
H_0 \cos\beta \sin\delta=0 \;.
}

\subsection{Branch I: solutions with $\alpha=0$}
In order to solve branch I, it is convenient to first examine a number of lower form conditions that follow from \eqref{eq: IIbSUSY}. To do this it is useful to first rotate the canonical frame of \eqref{eq: canonicalframe} such that
\eq{\label{eq: killframe}
v_1 &\to \sin\phi w_2+ \cos\phi(\cos\theta v_1+\sin\theta(\cos\delta w_1- \sin\delta v_2))\\[2mm]
v_2 &\to \cos\delta v_2+ \sin\delta w_1 \\[2mm]
w_1 &\to\cos\theta(\cos\delta w_1- \sin\delta v_2)- \sin\theta v_1\\[2mm]
w_2 &\to \cos\phi w_2-\sin\phi(\cos\theta v_1+ \sin\theta(\cos\delta w_1- \sin\delta v_2))\;.
}
Making use of these, one finds that the supersymmetry equations imply
\begin{subequations}
\begin{align}
&\sin\beta H_0=0~,\label{eq:simpcond1}\\[2mm]
&d(e^{2A+2C-\Phi} \cos\beta \cos\delta)- 2\nu e^{2A+ C-\Phi} \cos\beta v_2=0~,\label{eq:simpcond2}\\[2mm]
&d(e^{2A+3C-\Phi} \sin\beta(\cos\delta v_2+ \sin\delta w_1))- e^{2A+3C-\Phi} H_0 \cos\beta\cos\delta v_1\wedge w_2=0 ,\label{eq:simpcond3}\\[2mm]
& \cos\beta(e^C\cos\delta d\theta+2\nu \sin\delta v_1)=\cos\beta(e^{C}\cos\delta \sin\theta d\phi-2\nu \sin\delta w_2)=0~,\label{eq:simpcond4}\\[2mm]
& d(e^{3A+2C-\Phi} \sin\beta(\cos\delta w_1- \sin\delta v_2 ))\!+\! 2\nu e^{3A+C-\Phi}(\sin\beta v_2\wedge w_1+ \sin\delta \cos\beta v_1\wedge w_2)\nn\\[2mm]
&+e^{3A+2C-\Phi}\cos\beta \cos\delta (d\theta \wedge w_2+ \sin\theta d\phi\wedge v_1)=0,\label{eq:simpcond5}
\end{align}
\end{subequations}
which is not a compete list. The first thing to establish is how to solve \eqref{eq:simpcond1} - if we set $\sin\beta=0$, one needs to set $\cos\delta =0$ to solve  \eqref{eq:simpcond3}, but since $\nu = \pm 1$, \eqref{eq:simpcond2} leads to a contradiction.

The next conditions we consider are \eqref{eq:simpcond4}. For $\cos\beta\neq 0$ we see that either $\sin\delta=d\theta=d\phi=0$, or  $0<\sin\delta<\frac{\pi}{2}$ in which case $(\theta,\phi)$ define local coordinates on a 2-sphere. We are ignoring $\cos\beta=0$ because, as should be clear from, \eqref{eq: branchoneparameterisation}, this is a subcase of $\sin\delta=d\theta=d\phi=0$. Let us now prove that $0<\sin\delta<\frac{\pi}{2}$ is not possible: Since $H_0=0$ we can solve \eqref{eq:simpcond2}-\eqref{eq:simpcond3} by introducing local coordinates $x$ and $\rho=e^{2A+2C-\Phi} \cos\beta \cos\delta$ such that
\beq
\nu v_2=\frac{1}{2}\sqrt{\frac{\cos\delta}{\rho\cos\beta}}e^{-A+\frac{\Phi}{2}} d\rho,~~~~w_1=\frac{\sqrt{\cos\beta} \cos^{3/2}\delta}{2\nu \rho^{3/2}\sin\delta}e^{A-\frac{\Phi}{2}}(2 \nu \cot\beta dx- \sec\beta e^{-2A+\Phi}\rho d\rho).\nn
\eeq
We can also rewrite \eqref{eq:simpcond5} as
\beq
 d(e^{3A+2C-\Phi} \sin\beta(\cos\delta w_1- \sin\delta v_2 ))\!+\! 2\nu e^{3A+C-\Phi}(\sin\beta v_2\wedge w_1- \sin\delta \cos\beta v_1\wedge w_2)=0,\nn
\eeq
using \eqref{eq:simpcond4}. The key point here is that $v_2,~w_1$ only have legs in $(\rho,~x)$ while $v_1,~w_2$ sit orthogonal to this with legs in $(\theta,~\phi)$ only.  This means that the equation above, cannot be solved as there is a $\text{Vol}(S^2)$ term whose coefficient is non-vanishing. Thus we can conclude in general that $\sin\delta= d\theta=d\phi=0$. Plugging this back into \eqref{eq: IIaSUSY}, one finds that nothing depends on the specific values these paramaters take so we can set
\beq
H_0=\theta= \phi= \delta=0~.
\eeq
without loss of generality, leaving one undetermined function $\beta$.

We are now ready to write the supersymemtry conditions that follow when $\alpha=0$, however we find it helpful to perform a second rotation of the canonical vielbein by considering
\beq\label{eq:branchIsimplerot}
v_1+ i w_2 \to e^{-i \beta}w,~~~~ v_2-i w_1\to -i v,
\eeq
to ease presentation. The necessary and sufficient conditions for supersymmetry in the $\alpha=0$ branch are
\eq{
&H_0=d(e^{2A-\Phi} \sin\beta)= d(e^{A+C-\frac{1}{2}\Phi} \sqrt{\cos\beta})- \nu e^{A-\frac{\Phi}{2}} \sqrt{\cos\beta} v_2= d(e^{2A+3C-\Phi}\sin\beta v_2)=0~,\\[2mm]
& d(e^{3A+2C-\Phi} w)+ 2\nu e^{3A+C-\Phi}  w\wedge v_2=d(e^{3A+2C-\Phi}\sin\beta v_1)+ 2\nu e^{3A+C-\Phi} \sin\beta v_1\wedge v_2=0~,\\[2mm]
&d(e^{2A+2C-\Phi} v_1\wedge w)-2 \nu e^{2A+C-\Phi}v_1\wedge w\wedge v_2=0~,\\[2mm]
&d(e^{2A+2C-\Phi}\sin\beta w_1\wedge w_2)-2 \nu e^{2A+C-\Phi}\sin\beta w_1\wedge w_2\wedge v_2+ e^{2A+2C-\Phi} \cos\beta H_3=0~,\\[2mm]
&H_3+2\beta\wedge w_1\wedge w_2=d(e^{4A-\Phi} \cos\beta)\wedge w_1\wedge w_2\wedge v_2=0~,\\[2mm]
&d(e^{3A-\Phi}\cos\beta v_1)-d(e^{3A-\Phi}\sin\beta v_1\wedge w_1\wedge w_2)+ e^{3A-\Phi}\cos\beta v_1\wedge H_3-e^{3A}\star_4 \lambda(G_+)=0~,\\[2mm]
&d(e^{3A+3C-\Phi}\cos\beta v_1\wedge v_2)-e^{3C+3A}\star_4\lambda( G_-)=0~,\label{eq: alphazerobranchsusyIIB}\\[2mm]
&\bigg((\sin\beta +\cos\beta w_1\wedge w_2)\wedge v_2\wedge\lambda(G_-)-(\sin\beta+ \cos\beta w_1\wedge w_2)\wedge \lambda(G_+) \bigg)\bigg\lvert_{4}=0 .
}
We can simplify this system further, but not without making assumptions about $\beta$. We now proceed to study the systems that follow from different values of $\beta$, we find that the physical interpretation is quite different in each case.

\subsubsection{Sub case: $\beta=0$}\label{sec:first}
Upon setting $\beta=0$ in \eqref{eq: alphazerobranchsusyIIB} one can show that the supersymmetry conditions reduce to
\begin{subequations}
\begin{align}
&H_3 = H_0=d(e^{-\Phi})\wedge w_1\wedge w_2=d(e^{-4A})\wedge v_2\wedge w_1\wedge w_2=0~,\label{eq: SUSYIIB1b}\\[2mm]
&d(e^{-A}v_1)\wedge w=d(e^{A+C-\frac{1}{2}\Phi})- \nu e^{A-\frac{1}{2}\Phi}v_2= d(e^{A}w)=0~,\label{eq: SUSYIIB1a}\\[2mm]
&e^{3A}\star_4 \lambda(G_+)= d(e^{3A-\Phi}v_1),\quad e^{3A+3C}\star_4 \lambda(G_-)=  d(e^{3A+3C-\Phi}v_1\wedge v_2)\label{eq: SUSYIIB1c},\\[2mm]
& \lambda(G_-)\wedge v_2\wedge w_1\wedge w_2-\lambda(G_+)\wedge w_1\wedge w_2=0\label{eq: SUSYIIB1d}.
\end{align}
\end{subequations}
We can solve \eqref{eq: SUSYIIB1a} by using it to define a vielbein in terms of local coordinates $\psi,x_1,x_2$ and
\beq
\rho= e^{A+C-\frac{1}{2}\Phi}
\eeq
such that
\eq{\label{eq:vielbeinIIB1}
v_1&= e^{A}(d\psi+V)~,~~~~ v_2= \nu e^{-A+\frac{1}{2}\Phi}d\rho~,~~ w= e^{-A}(dx_1+ i dx_2)~,\\
V&=f_1(x_1,x_2)dx_1+ f_2(x_1,x_2)dx_2 \;.
}
From \eqref{eq: SUSYIIB1b} we see there is no NSNS flux, $\partial_{\psi}$ is an isometry and $A=A(\rho,x_1,x_2)$, $\Phi= \Phi(x_1,x_2)$. We now have enough information to calculate the fluxes. First we find
\eq{
\star_4 G_- &= - e^A \left(\p_{x_1} (e^{- \Phi})  w_1  + \p_{x_2} (e^{- \Phi}) w_2 \right) \wedge v_1 \wedge v_2
              - e^{3A - \Phi} \left( \p_{x_1} f_2 - \p_{x_2} f_1 \right) v_1 \\
\star_4 G_+ &= - e^{3A} \left(
 \p_{x_1} (e^{4A - \Phi}) w_1 \wedge v_2 + \p_{x_2} (e^{4A - \Phi})   w_2 \wedge v_2     + \n e^{- \frac12 \Phi} \p_\rho e^{4A- \Phi} v_2 \wedge v_1 \right) \\
&\phantom{=}- e^{3A - \Phi} \left(\p_1 f_2 - \p_2 f_1 \right) w_1 \wedge w_2 ~.
}
We can then use coordinate dependence of the physical fields and local expression for the vielbein \eqref{eq:vielbeinIIB1} to take the Hodge dual in \eqref{eq: SUSYIIB1c} arriving at
\begin{align}
G_- &= (\partial_{x_2}f_1-\partial_{x_1}f_2)e^{4A-\Phi}(d\psi+ V)+ \partial_{x_2}(e^{-\Phi})dx_1- \partial_{x_1}(e^{-\Phi})dx_2, \\[2mm]
G_+ &=-\nu e^{3A-\frac{3}{2}\Phi}\bigg( \partial_{x_2}(e^{-4A+ \Phi}) dx_1\wedge d\rho-\partial_{x_1}(e^{-4A+ \Phi}) dx_2\wedge d\rho-e^{-\Phi}\partial_{\rho}(e^{-4A+ \Phi}) dx_1\wedge dx_2\nn\\[2mm]
&-\nu e^{\Phi}(\partial_{x_2}f_1-\partial_{x_1}f_2) (d\psi+ V)\wedge d\rho\bigg).
\end{align}
Plugging this into \eqref{eq: SUSYIIB1d} we find $(\partial_{x_2}f_1-\partial_{x_1}f_2)=0$ which mean that $V$ is closed and so we can locally fix
\beq
V=0~,
\eeq
with a shift $\psi\to \psi - \eta$ for $d\eta=V$, without loss of generality. Taking this into account the ten-dimensional fluxes are
\begin{align}\label{d3d72}
F_1&=\partial_{x_2}(e^{-\Phi})dx_1- \partial_{x_1}(e^{-\Phi})dx_2,~~~~F_5=d\psi\wedge d(e^{4A-\Phi})\wedge\text{Vol}_3 \\[2mm]
&-\nu\rho^3\bigg( \partial_{x_2}(e^{-4A+ \Phi}) dx_1\wedge d\rho-\partial_{x_1}(e^{-4A+ \Phi}) dx_2\wedge d\rho-e^{-\Phi}\partial_{\rho}(e^{-4A+ \Phi}) dx_1\wedge dx_2\bigg)\wedge\text{Vol}(S^3).\nn
\end{align}
The final thing we need to do is impose the Bianchi identities, which away from localised sources rise to the PDEs
\eq{\label{d3d73}
(\partial_{x_1}^2+\partial_{x_2}^2)e^{-\Phi}=0~,~~~
\frac{e^{-\Phi}}{\rho^3}\partial_{\rho}(\rho^3 \partial_{\rho}e^{-4A+\Phi})+(\partial_{x_1}^2+\partial_{x_2}^2)( e^{-4A+\Phi})=0~.
}
The local form of the metric is then
\eq{\label{d3d71}
ds^2 &=\frac{1}{\sqrt{f H}} ds^2(\mathbb{R}^{1,3})+ \sqrt{\frac{H}{f}} \bigg(d\rho^2+ \rho^2 ds^2(S^3)\bigg)+ \sqrt{f H}\left(dx_1^2+ dx_2^2\right)~,\\
H &= e^{-4 A + \Phi}~,~~~f = e^{- \Phi} ~
}
This corresponds to the intersecting D3-D7 brane system, where the D3-branes are embedded in the D7-branes \cite{youm}.

\subsubsection{Sub case: $\beta=\frac{\pi}{2}$}\label{eq:D5case}
Setting $\beta=\frac{\pi}{2}$ in \label{eq: alphazerobranchsusyIIB} leads to the following nessesary and sufficient condtions for unbroken supersymmetry
\begin{subequations}
\begin{align}
&d(e^{2A-\Phi})= d(e^{2A+ 3C-\Phi} v_2)= H_3=H_0=G_-=0~,\label{eq:D5casebps1a}\\[2mm]
&d( e^{3A+2C-\Phi}u_i)+ 2\nu e^{3A+C-\Phi} u_i\wedge v_2=0~,\label{eq:D5casebps1b}\\[2mm]
&d(e^{2A+2C-\Phi} \epsilon_{ijk} u_j\wedge u_k)- 2\nu e^{2A+C-\Phi}\epsilon_{ijk}u_j\wedge u_k\wedge v_2=0~,\label{eq:D5casebps1c}\\[2mm]
&e^{3A}\star_4 \lambda(G_+)= d(e^{3A-\Phi} v_1\wedge w_1\wedge w_2)\label{eq:D5casebps1d}.
\end{align}
\end{subequations}
Here we have introduced the notation
\beq
u=(v_1,w_1,w_2),
\eeq
both to ease notation and to stress that the vielbeine $u_i$ obey a cyclic property. Exploiting this property will be very helpful in solving this system and other systems we shall encounter which mirror this behaviour, so we will be very explicit in our derivation here, but less so elsewhere.  The first thing to note is that the combination  \eqref{eq:D5casebps1c}$_i+ e^{-A}\epsilon_{ijk} u_j\wedge$\eqref{eq:D5casebps1b}$_k$ leads to
\beq
\epsilon_{ijk} \bigg(d(e^{2A+C-\frac{1}{2}\Phi})-\nu e^{2A-\frac{1}{2}\Phi}v_2\bigg)\wedge u_j\wedge u_k=0~.
\eeq
This implies that the 1-form in large brackets is zero. This can be seen by writing it as $X_j u^j + X_0 v_2$ for some functions $X_j$, noting that the vielbeine $u_{1,2,3}$ are independent, and then considering the resulting constraints for $i=1,2,3$. Next by examining  \eqref{eq:D5casebps1c}$_i+ e^{-A} u_j\wedge$\eqref{eq:D5casebps1b}$_k$ and \eqref{eq:D5casebps1c}$_i- e^{-A}\wedge$\eqref{eq:D5casebps1b}$_j\wedge u_k$ for cyclic permutations of $(i,j,k)=(1,2,3)$ one realises that
\beq
d(e^{-A}u_i)\wedge u_j =0~,~~~~ i\neq j,
\eeq
which implies that $d(e^{-A}u_i)$ has no leg in $u_i$, it is then not hard to see that since \eqref{eq:D5casebps1b} has no $\epsilon_{ijk} u_j\wedge u_k$ term it is in fact zero. So we can conclude without loss of generality that
\beq
d(e^{-A} u_i)= d(e^{2A+C-\frac{1}{2}\Phi})-\nu e^{2A-\frac{1}{2}\Phi}v_2=0~,
\eeq
which imply \eqref{eq:D5casebps1b}-\eqref{eq:D5casebps1c} without further constants. We can then solve these conditions by using them to define the vielbeine in terms of local coordinates as
\beq
v_1 = e^A d x_1~,~~~w_1 =  e^A  d x_2~,~~~w_2 = e^A d x_3~,~~~v_2 = \nu e^{-2A+\frac{1}{2}\Phi} d\rho~,~~~ \rho=e^{2A+C-\frac{1}{2}\Phi}~.
\eeq
We can now solve \eqref{eq:D5casebps1a}, which in fact just tells us that
\beq
e^{\Phi}= e^{2A},
\eeq
up to rescaling $g_s$ and that $e^{2A}$ is a function of $\rho$ only, making $\partial_{x_i}$  all isometry directions parameterising either $\mathbb{R}^3$ or $T^3$ locally.

The only non-trivial flux is the RR 3-form
\beq\label{d52}
F_3= - \nu \rho^3 \partial_{\rho}( e^{-4A}) \text{Vol}(S^3)
\eeq
and its Bianchi identity, $dF_3=0$, imposes that
\beq\label{d53}
e^{-4A}= c_1 + \frac{c_2}{\rho^2},~~~ dc_i=0~.
\eeq
This is the warp factor of a D5-brane or O5-hole, depending on the sign of $c_2$ (see for example \cite{johnson}). Indeed the metric locally takes the form
\beq\label{d51}
ds^2= e^{2A}ds^2(\mathbb{R}^{1,5})+  e^{-2A}\bigg(d\rho^2+ \rho^2ds^2(S^3)\bigg)
\eeq
As we will see, this is a subcase of the solution in the next section.

\subsubsection{Sub case: Generic $\beta$}
For generic $0<\beta<\frac{\pi}{2}$  we are free to divide by the trigonometric functions in \eqref{eq: alphazerobranchsusyIIB}. Using $\sin\beta \neq 0$  it is possible to show that supersymmetry requires
\begin{subequations}
\begin{align}
&d(e^{2A-\Phi}\sin\beta)=d(e^{-\Phi}\cos\beta)=d e^{A}\wedge v_2=0 ,\label{eq:SUSYIIB3a},\\[2mm]
&d(e^{-A}v_1)= d(e^{2A+C-\frac{\Phi}{2}}\sqrt{\sin\beta})-\nu e^{2A-\frac{\Phi}{2}}\sqrt{\sin\beta}v_2 = d(e^{-A}w \csc \beta)=0~,\label{eq: SUSYIIB3b}\\[2mm]
&H_3 + 2d\beta\wedge w_1\wedge w_2= d\beta\wedge v_2\wedge w_1\wedge w_2 =0~,\label{eq: SUSYIIB3c}\\[2mm]
&e^{3A}\star_4 \lambda(G_+)= d(e^{3A-\Phi}\cos\beta v_1)+d(e^{3A-\Phi}\sin\beta v_1\wedge w_1\wedge w_2)- e^{3A-\Phi}\cos\beta H_3\wedge v_1 ,\nn\\[2mm]
&e^{3A+3C}\star_4 \lambda(G_-)= d(e^{3A+3C-\Phi}\cos\beta v_1\wedge v_2)\label{eq: SUSYIIB3e},\\[2mm]
& \lambda(G_-)\wedge v_2\wedge(\sin\alpha+ \cos\alpha w_1\wedge w_2)- \lambda(G_+)\wedge(\sin\alpha+
 \cos\alpha w_1\wedge w_2)=0~.\label{eq: SUSYIIB3f}
\end{align}
\end{subequations}
by following the same line of reasoning as in the previous subsection. First we solve \eqref{eq: SUSYIIB3b} by using it to define the vielbeine on $M_4$ locally
\beq\label{eq:branch1vielIIBbeta}
v_1=e^{A}dx_1,~~~ w = e^{A}\sin\beta( dx_2+i dx_3),~~~ v_2= \nu e^{-A}d\rho,~~~ \rho= e^{A+C} \;,
\eeq
where we have used the first of \eqref{eq:SUSYIIB3a} to simplify these somewhat. Next \eqref{eq:SUSYIIB3a} is solved when
\beq
e^{\Phi}= e^{2A}\sin\beta,~~~ \cot\beta= c e^{2A},~~~dc=0~,
\eeq
with $A = A(\rho)$, $\beta = \beta (\rho)$ . As a result, $\partial_{x_i}$ are isometries. We then use \eqref{eq:branch1vielIIBbeta} to take the Hodge dual of \eqref{eq: SUSYIIB3e}, \eqref{eq: SUSYIIB3f} arriving at the fluxes
\begin{align}\label{d5u2}
F_3&=-\nu \rho^3\partial_{\rho}(e^{-4A})\text{Vol}(S^3),~~~ H= 2 e^{2A}\partial_{\rho}\beta\sin^2\beta d\rho\wedge dx_2\wedge dx_3,\\[2mm]
F_5&= \text{Vol}_3\wedge dx_1\wedge d(e^{2A} \cot\beta)+ \nu e^{-2A}\rho^3\left(\sin(2\beta) \partial_{\rho}A-\partial_{\rho}\beta\right)dx_2\wedge dx_3\wedge \text{Vol}(S^3)\nn,
\end{align}
which solve \eqref{eq: SUSYIIB3f} without restriction.
The Bianchi identities impose that
\beq\label{d5u3}
e^{-4A}= c_1+ \frac{c_2}{\rho^2}~,~~~dc_i=0~,
\eeq
which is again the warp factor of a D5-brane or O5-hole. However, in this case the metric takes the local form
\beq\label{d5u1}
ds^2= e^{2A} ds^2(\mathbb{R}^{1,3})+ e^{2A}\sin\beta^2 ds^2(T^2)+ e^{-2A}\bigg(d\rho^2+ \rho^2 ds^2(S^3)\bigg)
\eeq
where $T^2$ is spanned by $(x_2,x_3)$. This generalises the solution in the previous section by introducing an additional warping factor for a $T^2$ submanifold, thus breaking $SO(1,5)$ Lorentz symmetry and leading to more general fluxes. In fact, this solution can be generated from the D5-brane solution of the previous section via "$G$-structure rotation" \cite{Gaillard:2010qg} which is formally a U-duality \cite{Maldacena:2009mw}.

\subsection{Branch II: $\alpha$ non-zero solutions}
For the second branch with $0<\alpha<\pi$, we begin by studying the lower form conditions that follow from \eqref{eq: IIbSUSY}. Here we find it useful to rotate the canonical frame of \eqref{eq: canonicalframe} as
\begin{align}\label{eq: branchIIrot}
v&\to-\sin w_1+ \cos\phi(\cos\theta v_1- \sin\theta w_2)+ i v_2,\\[2mm]
w&\to \cos\phi w_1+\sin\phi (\cos\theta v_1- \sin\theta w_2)+ i (\cos\theta w_2 + \sin\theta v_1).
\end{align}
We then find the following necessary, but not sufficient, conditions for supersymmetry
\begin{subequations}
\begin{align}
&\cos\beta \sin\delta H_0=0~,\label{eq:branchIIIIBsimpa}\\[2mm]
& e^{C}\cos\alpha\sin\beta d\theta-2 \nu(\cos\beta \cos\delta v_1+ \sin\beta\sin\alpha w_1)=0~,\label{eq:branchIIIIBsimpb}\\[2mm]
&e^{C}\cos\alpha\sin\beta \sin\theta d\phi-2\nu(\cos\beta \cos\delta w_1- \sin\alpha \sin\beta v_1)=0~,\label{eq:branchIIIIBsimpc}\\[2mm]
&d(e^{2A+2C-\Phi} \cos\alpha \sin\beta)-2 \nu e^{2A+C-\Phi}(\cos\beta\cos\delta w_2- \sin\beta v_2),\label{eq:branchIIIIBsimpd}\\[2mm]
&d(e^{2A+3C-\Phi}(\sin\alpha \sin\beta w_2+ \sin(\alpha\!-\!\delta)\cos\beta v_2))+ e^{2A+3C-\Phi} H_0 \sin\beta \cos\alpha v_1\wedge w_1=0\label{eq:branchIIIIBsimpe},\\[2mm]
&d(e^{3A+2C-\Phi}\left(\sin\beta\sin\alpha v_2- \cos\beta\sin(\alpha-\delta )w_2 )\right)
+ 2\nu e^{3A+C-\Phi}\cos\beta(\sin\delta w_2\wedge v_2+ \cos(\alpha-\delta)v_1\wedge w_1)\nn\\[2mm]
&-e^{3A+2C-\Phi}\left(d\theta\wedge (\cos\beta\sin(\alpha\!-\!\delta)v_1+ \sin\beta w_1)+\sin\theta d\phi\wedge(\cos\beta\sin(\alpha\!-\!\delta) w_1-\sin\beta v_1 )\right)=0\label{eq:branchIIIIBsimpf}~.
\end{align}
\end{subequations}
First we note that if either $\theta$ or $\phi$ become constant or if $\cos\alpha=0$ then \eqref{eq:branchIIIIBsimpb}, \eqref{eq:branchIIIIBsimpc} require that $\sin\beta=\cos\delta=0$ which makes $(\theta,~\phi)$ drop out of \eqref{eq:branchIIparamaterisation} entirely and the final line of \eqref{eq:branchIIIIBsimpf} vanishes  (setting $\sin\theta=0$  leads to the same conclusion).
In this case we can conclude that we can set
\beq
H_0=\theta=\phi=(\delta-\frac{\pi}{2})=0
\eeq
without loss of generality, which we study in section \ref{sec: additional S3}.

If we assume $\sin\theta$ and $\cos\alpha$ don't vanish, then $(\theta,\phi)$ are local coordinates on a 2-sphere and we can take $\rho=e^{2A+2C-\Phi} \cos\alpha \sin\beta$ as a local coordinate. We can then use \eqref{eq:branchIIIIBsimpb}-\eqref{eq:branchIIIIBsimpd} to rewrite \eqref{eq:branchIIIIBsimpf} as
\beq
d(e^{3A+2C-\Phi}(\sin\beta\sin\alpha v_2- \cos\beta\sin(\alpha\!-\!\delta)w_2 ))+ 2\nu e^{3A+C-\Phi}\cos\beta(\sin\delta w_2\wedge v_2- \cos(\alpha\!-\!\delta)v_1\wedge w_1)=0,\label{eq:branchIIthing}
\eeq
which we can then use to fix some of the free functions. First we note that if we solve \eqref{eq:branchIIIIBsimpa} with $\sin\delta=0$, then \eqref{eq:branchIIthing} fixes $\cos\beta=0$ - this is because $\sin\beta\sin\alpha v_2- \cos\beta\sin\alpha w_2$ is parallel to $d\rho$ in this limit and so cannot generate the $\text{Vol}(S^2)$ factor that comes from $v_1\wedge w_1$. Next, for $H_0=0$ and for generic values of $(\alpha,\beta,\delta)$ we can use \eqref{eq:branchIIIIBsimpb}-\eqref{eq:branchIIIIBsimpe} to locally define the vielbein on $M_4$ by introducing another local coordinate such that $dx=e^{2A+3C-\Phi}(\sin\alpha \sin\beta w_2+ \sin\alpha\!-\!\delta\cos\beta v_2)$,  but then we must once more set the $v_1\wedge w_1$ term in \eqref{eq:branchIIthing} to zero which fixes either $\cos\beta=0$ or $\cos(\alpha-\delta)=0$. Thus, for $H_0=0$ and a priori generic $(\theta,\phi,\alpha,\beta,\delta)$, we end up with just two cases. Firstly $\cos\beta=0$, which solves \eqref{eq:branchIIIIBsimpa} and makes the $\delta$ dependence of \eqref{eq:branchIIparamaterisation} drop out such that we can set without loss of generality
\beq
\delta=0~,~~~ \beta=\frac{\pi}{2}~.
\eeq
 We shall examine this case in detail in section \ref{sec:complicated IIB} where we find that it contains no solution.
Secondly $\cos(\alpha-\delta)=0$, such that we can set without loss of generality
\beq
\delta=\alpha+\frac{\pi}{2}~,
\eeq
which we shall study in section \ref{sec:complicated 2 IIB}, finding a new class of solution.

There is one final option one can consider for $H_0=0$, by taking both $\cos\alpha$ and $\sin\theta$ non vanishing- one can tune the values of $(\alpha,\delta,\beta)$ such that $(\sin\alpha \sin\beta w_2+ \sin(\alpha\!-\!\delta)\cos\beta v_2))$ becomes parallel to  $d\rho$ and so can no longer be used to introduce a local coordinate. This requires fixing
\beq
\tan\beta =\sqrt{\frac{\cos\delta \sin(\delta-\alpha)}{\sin\alpha}}.
\eeq
It will then be \eqref{eq:branchIIthing} that will be used to define the final vielbein direction, which will necessarily be fibred over $S^2$. We shall examine this possibility in section \ref{sec:u1overs2}.

\subsubsection{Sub case $\beta=0$}\label{sec: additional S3}
For $\beta=0$ the supersymmetry conditions reduce to
\begin{subequations}
\begin{align}
&d(e^{2A-\Phi})= d(e^{2A+3C-\Phi}\cos\alpha v_2)=H_3=H_0=0~,\label{eq:additionalS3BPSa}\\[2mm]
&d(e^{3A+2C-\Phi}\cos\alpha u_i)+2\nu e^{3A+C-\Phi}(u_i\wedge v_2+ \frac{1}{2}\sin\alpha \epsilon_{ijk}u_j\wedge u_k)=0~,\label{eq:additionalS3BPSb}\\[2mm]
&d(e^{2A+2C-\Phi}(\sin\alpha u_i\wedge v_2+\frac{1}{2}\epsilon_{ijk}u_j\wedge u_k))-\nu e^{2A+C-\Phi}\cos\alpha  \epsilon_{ijk}u_j\wedge u_k\wedge v_2=0~,\label{eq:additionalS3BPSc}\\[2mm]
&\!e^{3A}\star_4\lambda(G_+)=\! d(e^{3A-\Phi} \cos \a v_1\wedge w_1\wedge w_2),~~e^{3A+3C}\star_4\!\lambda(G_-)=-d(e^{3A+3C-\Phi}\sin\alpha)\label{eq:additionalS3BPSd},\\[2mm]
&(\cos\alpha v_2\wedge\lambda(G_-)+\sin\alpha v_1\wedge v_2\wedge w_1 \wedge w_2\lambda(G_+))\big\lvert_4=0\label{eq:additionalS3BPSe}.
\end{align}
\end{subequations}
where as usual $u=(v_1,w_1,w_2)$. Note that this system reduces to that of section \ref{eq:D5case} when $\sin\alpha=0$, and that \eqref{eq:additionalS3BPSb} imposes that $\cos\alpha \neq 0$, so we can take $0<2\alpha<\pi$. By adding linear combinations of wedge products of \eqref{eq:additionalS3BPSa}, \eqref{eq:additionalS3BPSb} and the vielbein to \eqref{eq:additionalS3BPSb}, it is then possible to derive enough independent 2-form conditions to establish that
\beq\label{eq:IIBthing1}
d(e^{A-C}\sin\alpha)\wedge v_2=d\alpha \wedge v_2 = d(e^{A+C}\cos\alpha)- \nu e^{A} v_2 = d(e^{-A}\sec \alpha u_i)\wedge u_j=0, ~~~i \neq j.
\eeq
This is sufficient to establish that $e^{2A}$, $e^{2C}$ and $\alpha$  are functions of a single local  coordinate $\rho$, which $v_2$ is parallel to - specifically
\beq
\rho=e^{A+C}\cos\alpha~,~~~~ v_2 = \nu e^{-A}d\rho~.
\eeq
The final condition in \eqref{eq:IIBthing1} implies that  $d(e^{-A}\sec\alpha u_1) \propto u_2\wedge u_3$ and cyclic permutations, however plugging this into \eqref{eq:additionalS3BPSb} and \eqref{eq:additionalS3BPSb} we realise we can without loss of generality take
\beq
u_i= \frac{c_1}{2}e^{A}\cos\alpha \tilde{K}_i~,~~~\tilde{K}_i+\frac{\nu}{2}\epsilon_{ijk}\tilde{K}^j\wedge \tilde{K}^k~,~~~
c_1= \frac{e^{-A+C}}{\sin\alpha}~,~~~ dc_1=0~,
\eeq
where $\tilde{K}^i$ are necessarily $SU(2)$ invariant forms which furnish a frame for a round $S^3$. We have now without loss of generality determined the vielbein on $M_4$, which is a foliation of $S^3$ over an interval, and \eqref{eq:additionalS3BPSa}-\eqref{eq:additionalS3BPSc} are solved when
\beq
e^{-2A}=\frac{c_1 \sin\alpha \cos\alpha}{\rho}~,~~~ e^{2C}=e^{2A} c_1^2 \sin^2\alpha~,~~~ e^{2A-\Phi}=c_2~,~~~dc_i=0~.
\eeq
The only non-trivial 10d flux can be extracted from \eqref{eq:additionalS3BPSd} and is given by
\beq\label{d5s2}
F_3 = 2 c_1^2c_2\nu\big(\sin^2\alpha-\rho \tan\alpha\partial_{\rho}\alpha\big)\text{Vol}(S^3)+ 2 c_1^2c_2 \nu\big(\cos^2\alpha+\rho \cot\alpha\partial_{\rho}\alpha\big)\text{Vol}(\tilde S^3)~.
\eeq
The pairing equation \eqref{eq:additionalS3BPSe} is equivalent to the Bianchi identity at this point; either one implies
\beq\label{d5s3}
 d \a = 0~.
\eeq
The metric is of the form
\begin{align}\label{eq: d5thing}
ds^2&= e^{2A}ds^2(\mathbb{R}^{1,2})+e^{-2A}\bigg[d\rho^2+\frac{\rho^2}{\cos^2\alpha} ds^2(S^3)+ \frac{\rho^2 }{\sin^2\alpha }ds^2(S^3)\bigg]
\end{align}
This solution has both an $SO(4)$ R-symmetry and $SO(4)$ flavour symmetry, and is S-dual to the one that we find in section \ref{e20}, as will be explained in that section.

\subsubsection{Sub case $\beta=\frac{\pi}{2}$}\label{sec:complicated IIB}
Here one can show that supersymmetry implies
\begin{subequations}
\begin{align}\label{nope0}
&d(e^{2A+2C-\Phi}\cos\alpha)-2\nu e^{2A+C-\Phi}v_2= d(e^{C}\cot\alpha w_2)=0~,\\[2mm]
&e^{C}d\theta-2\nu \tan\alpha w_1= e^{C}\sin\theta d\phi+2\nu \tan\alpha v_1=0~,\\[2mm]
&d(e^{-A-C}\sin\alpha)= d(e^{\Phi}\sin\alpha\tan\alpha)\wedge v_2= d(e^{-A-C+\Phi}\tan\alpha)\wedge w_2\wedge v_2=0~,\\[2mm]\label{nope1}
&d(e^{A+C-\frac{\Phi}{2}}\tan\alpha\sqrt{\cos\alpha})\wedge w_2+ \frac{1}{2}\sqrt{\cos\alpha}e^{A+C-\frac{\Phi}{2}}H_0 v_1\wedge w_1=0~,\\[2mm]\label{nope2}
&d(e^{2A+2C-\Phi}\cos\alpha \cot^2\alpha)\wedge \text{Vol}(S^2)=0~,\\[2mm]
&H_3+ \frac{\nu}{2} e^{C}\cot\alpha w_2\wedge \text{Vol}(S^2)= d(e^{-2C}\tan\alpha)\wedge v_2\wedge \text{Vol}(S^2)=0~,\\[2mm]
&e^{3A}\star_4\lambda(G_+)-d(e^{3A-\Phi}w_2)+ d(e^{3A-\Phi}\sin\alpha v_1\wedge v_2\wedge w_1)=0~,\\[2mm]
&e^{3A+3C}\star_4\lambda(G_-)+ e^{3A+3C-\Phi}H_0 w_2+ d(e^{3A+3C-\Phi}\cos\alpha v_2\wedge w_2)=0~,\\[2mm]
&\bigg((\sin\alpha w_2- v_1\wedge v_2\wedge w_1)\wedge \lambda(G_-)- \cos\alpha v_1\wedge v_2\wedge \lambda(G_+)\bigg)\bigg\lvert_{4}=0~.
\end{align}
\end{subequations}
There is no solution to this set of constraints, as we will now show. First, we note that $H_0 = 0$ is imposed by \eqref{nope1}. Let $\rho = e^{2A + 2C - \Phi} \cos \a$. Due to \eqref{nope0}, we have $w_2 \sim d x$, $v_2 \sim d \rho$. It is then possible to rewrite \eqref{nope1}, \eqref{nope2} as
\eq{
d \left( \sqrt{\rho} \tan \a \right) \wedge d x &= 0~,~~~~
d \left( \rho \cot^2 \a \right) \wedge d \t \wedge d \phi = 0 ~.
}
The first equation implies $\tan \a = \rho^{-1/2} f(x)$, which is incompatible with the second equation - thus this putative class contains no solutions.

\subsubsection{Sub case $\delta=\alpha+\frac{\pi}{2}$, $H_0=0$}\label{sec:complicated 2 IIB}
As explained below \eqref{eq:branchIIIIBsimpf}, here we necessarily have $\beta>0$ and $0<\alpha<\frac{\pi}{2}$  - for this reason it will turn out the case contains no solutions. As the proof is similar to that of the previous section we shall be brief, this time only quoting sufficient supersymmetry conditions to prove this. In addition to the rotation of \eqref{eq: canonicalframe} we find it useful to send $v_1+ i w_1 \to e^{-i\beta}(v_1+ i w_1)$, then a set necessary (but insufficient) conditions for supersymmetry are
\begin{subequations}
\begin{align}
&d(e^{2A-\Phi}\cos\alpha \cos\beta)=0,\label{eq:nope2a1}\\[2mm]
& d(e^{C}\cot\alpha w_2)=0,\label{eq:nope2a2} \\[2mm]
&(v_1+i w_1)\tan\alpha+ \frac{e^{C}\sin\beta}{2\nu} (d\theta+i \sin\theta d\phi),~~~d(e^{2A+3C-\Phi}(\sin\alpha\sin\beta w_2- \cos\beta v_2))=0,\label{eq:nope2b}\\[2mm]
&d(e^{2A+2C-\Phi}\cos\alpha \sin\beta)-2 \nu e^{2A+C-\Phi}(\cos\beta\sin\alpha w_2+ \sin\beta v_2)=0,\label{eq:nope2c} \\[2mm]
&e^{2A-\Phi}\cos\alpha \cos\beta H_3+ d(e^{2A-\Phi}\cos\alpha \sin\beta v_1\wedge w_1)=0,\label{eq:nope2d} \\[2mm]
&e^{2A+3C-\Phi}(\sin\alpha \sin\beta w_2- \cos\beta v_2)\wedge H_3+ d(e^{2A+3C-\Phi} v_1\wedge w_1 \wedge(\sin\beta v_2- \cos\beta \sin\alpha w_2))=0.\label{eq:nope2e}
\end{align}
\end{subequations}
As elsewhere we can take \eqref{eq:nope2b}-\eqref{eq:nope2c} as a local definition of the vielbein without loss of generality - $v_1,w_1$ are clearly the local vielbeine of a round $S^2$ in terms of the local coordinates $(\theta,~\phi)$. For $v_2,~w_2$ we introduce local coordinates $x$ and $\rho=e^{A+C-\frac{1}{2}\Phi}\sqrt{\cos\alpha \sin\beta}$ such that
\beq
e^{2A+3C-\Phi}(\sin\alpha\sin\beta w_2- \cos\beta v_2)=dx.
\eeq
We can then use \eqref{eq:nope2d} to define $H_3$ without loss of generality, which leaves \eqref{eq:nope2a1}, \eqref{eq:nope2a2} and \eqref{eq:nope2e} to solve - this turns out to be impossible. To see this, one needs to consider the combination $4$\eqref{eq:nope2e}$+ \big(f_1 $\eqref{eq:nope2a1}$\wedge d\rho+f_2 $\eqref{eq:nope2a1}$\big)\wedge \text{Vol}(S^2)$. When one tunes
\beq
f_1=e^{-A+5C+\frac{\Phi}{2}} \nu\frac{ \cos^{\frac{3}{2}}\alpha\sin^{\frac{5}{2}}\beta}{\sin^2\alpha \cos\beta},~~~f_2= e^{2A+4C-\Phi}\cos\alpha \sin\beta\tan\beta,
\eeq
this leads to
\beq
\cot\alpha \csc\alpha \sin\beta \tan\beta dx\wedge d\rho\wedge\text{Vol}(S^2)=0
\eeq
which cannot be solved without violating the the initial assumptions that lead to this case. We conclude that there exist no solutions.

\subsubsection{Sub case: Special value of $\beta$, $H_0=0$}\label{sec:u1overs2}
The final case in IIB requires us to tune $\tan\beta$ to a specific value. After redefining $\delta\to\delta+\alpha$, this value is
\beq
\tan\beta=\sqrt{\frac{\cos(\alpha+\delta) \sin\delta}{\sin\alpha}}~.
\eeq
In addition, we rotate the vielbein (with respect to \eqref{eq: branchIIrot}) as
\begin{align}
  v_1+i w_1 &\to -i \frac{\cos\beta \cos(\alpha+\delta)+ i \sin\alpha \sin\beta}{\sqrt{\cos^2\beta \cos^2(\alpha+\delta)+\sin^2\alpha \sin^2\beta}}(v_1+i w_1) \nn,\\[2mm]
	  v_2+i w_2 &\to -i\frac{\sin\alpha \sin\beta + i \cos\beta \sin\delta}{\sqrt{\sin^2\alpha\sin^2\beta+ \cos^2\beta \sin^2\delta}}( v_2+i w_2 )
\end{align}

In what follows we assume that the undefined functions of the spinor ansatz are bounded as $0<2\alpha<\pi$ and $0<\delta+\alpha< \frac{\pi}{2}$, as the upper and lower limits have been dealt with in the proceeding sections. It is then possible to show that the necessary and sufficient conditions for supersymmetry for this case are are
\begin{subequations}
\begin{align}
&d(e^{A-C}\sin\alpha)= d\left(e^{2A-\Phi}\sqrt{\frac{\sin\alpha \sin(\alpha+\delta)}{\cos\delta}}\right)=d\left(e^{-\Phi}\sqrt{\frac{\sin 2(\alpha+\delta)}{\sin 2\delta}}\right)=0,\label{eq:IIBbpsfinala}\\[2mm]
&2\nu e^{-C}\sqrt{\tan\alpha \cot\delta}(w_1-i v_1)- (d\theta+ i \sin\theta d\phi)=0,\label{eq:IIBbpsfinalb}\\[2mm]
&2\nu d\left(\!e^{-C}\tan\alpha\sqrt{\frac{\cos\alpha \sin(\alpha+\delta)}{\sin\delta}}w_2\right)-\text{Vol}(S^2)=d\left(\!e^{A+C}\sqrt{\frac{\cos\alpha \sin\delta}{\sin(\alpha+\delta)}}\right)-\nu e^{A}v_2=0,\label{eq:IIBbpsfinalc}\\[2mm]
&H_3+\frac{1}{4}d\left(e^{2C}\frac{\cos^2\alpha \sin\delta}{\sin^2\alpha\sin(\alpha+\delta)\cos\delta} \sqrt{\cos(\alpha+\delta)\sin\alpha\sin\delta }\right)\wedge \text{Vol}(S^2)=d\alpha\wedge v_2=0\label{eq:IIBbpsfinald}\\[2mm]
& e^{3A}\star_4\lambda(G_+)-d_{H_3}\left(e^{3A-\Phi}\sqrt{\frac{\cos\alpha \cos(\alpha+\delta)}{\cos\delta}}\right)-d\left(e^{3A-\Phi} \sqrt{\frac{\sin\alpha\cos\alpha\sin\delta }{\cos\delta}}v_1\wedge w_1\wedge w_2\right)=0,\label{eq:IIBbpsfinale}\\[2mm]
& e^{3A+3C}\star_4\lambda(G_-)+ d_{H_3} \left(e^{3A+3C-\Phi}\sqrt{\frac{\cos\delta \sin\alpha}{\sin(\alpha+\delta)}}\right)- d\left(e^{3A+3C-\Phi}\cos\alpha \sqrt{\frac{\sin\delta \cos(\alpha+\delta)}{\cos\delta \sin(\alpha+\delta)}}v_2\wedge w_2\right)=0,\nn\\[2mm]
&\bigg(\big(\sqrt{\frac{\sin\alpha\cos\alpha\sin\delta}{\cos\delta}}v_2- \sqrt{\frac{\cos\alpha\cos(\alpha+\delta)}{\cos\delta}}v_1\wedge w_1\wedge v_2\big)\wedge \lambda(G_-)\label{eq:IIBbpsfinalf}\\[2mm]
&- \big(\sqrt{\frac{\sin\alpha \sin(\alpha+\delta)}{\cos\delta}}-\cos\alpha\sqrt{\frac{\sin\delta \cos(\alpha+\delta)}{\cos\delta \sin(\alpha+\delta)}}v_1\wedge w_1- \sqrt{\frac{\sin\alpha\cos\delta}{\sin(\alpha+\delta)}}v_1\wedge v_2\wedge w_1\wedge w_2\big)\wedge \lambda(G_+)\bigg)\bigg\lvert_{4}=0\nn,
\end{align}
\end{subequations}
where $\text{Vol}(S^2)$  is the volume form on the $S^2$ spanned by $(\theta,~\phi)$. We solve these conditions by first using \eqref{eq:IIBbpsfinalb}-\eqref{eq:IIBbpsfinalc} to define the vielbein locally without loss of generality as
\begin{align}\label{eq:lastIIBvielbein}
w_1&= \frac{\nu}{2}e^{C}\sqrt{\frac{\cos\alpha\sin\delta}{\sin\alpha\cos\delta}}d\theta~,~~~
v_1=-\frac{\nu}{2}e^{C}\sqrt{\frac{\cos\alpha\sin\delta}{\sin\alpha \cos\delta}}\sin\theta d\phi~,\\[2mm]
w_2&=-\frac{\nu}{2}e^{C}\cot\alpha\sqrt{\frac{\sin\delta}{\sin(\alpha+\delta) \cos\alpha}}(d\psi+ \cos\theta d\phi)~,~~~
v_2= \nu e^{-A}d\rho~,
\end{align}
where we have taken $\theta,\phi$ as local coordinates coordinates and introduced the additional coordinates $\psi$ and
\beq
\rho = e^{A+C}\sqrt{\frac{\cos\alpha \sin\delta}{\sin(\alpha+\delta)}}~.
\eeq
We can invert this conditions then use  \eqref{eq:IIBbpsfinala} to define $A,C,\Phi,\delta$ in terms of $\alpha,\rho$ and some integration constants $c_i$ as
\begin{align}\label{eqoook}
e^{-4A}&=\frac{c_1^2 \cos^2\alpha \sin^2\alpha}{\rho^2} - \frac{c_3 \sin^2\alpha}{c_2^2},~~~ e^{2C}=c_1^2\sin^2\alpha e^{2A},\nn\\[2mm]
e^{-2\Phi}&= e^{-4A}\left(c_2^2+ \frac{c_3 \rho^2}{c_1^2 \sin\alpha^2}\right),~~~\cot(\alpha+\delta)= \frac{c_3 \rho^2}{c_1^2 c_2^2 \sin\alpha \cos\alpha}~.
\end{align}
The second equality in \eqref{eq:IIBbpsfinald} implies that $\alpha$ is itself a function of $\rho$ only, so we realise that $\partial_{\psi}$ is an isometry of the solution and that $M_4$ is foliation of  a ($SU(2)\times U(1)$ preserving) squashed 3-sphere over an interval.

We now turn our attention to the fluxes. We have that \eqref{eq:IIBbpsfinald} simply defines the NSNS flux in such a way that it is automatically closed, while the RR fluxes are defined through the 4d fluxes that follow from \eqref{eq:IIBbpsfinale}. We could use the definitions of the functions in \eqref{eqoook} and vielbein in \eqref{eq:lastIIBvielbein} to calculate the 10d fluxes immediately, however we already have enough information to  first fix $\alpha$. The 3-form component of $\star_4\lambda(G_-)$ is necessarily parallel to $v_1\wedge w_1\wedge v_2$ from which it follows that the 10d  flux $F_1$ is parallel to $w_2$. As this vielbein is fibred over the $S^2$ we have that $dF_1=0$ iff $(\star_4\lambda(G_-))_3=0$. For $\cos(\alpha+\delta)\neq 0$ this imposes $d(c_1^2 c_2^2\sin^2\alpha+ c_3 \rho^2)=0$, which implies that
\beq
\sin^2\alpha= c_4-\frac{c_3 \rho^2}{c_1^2c_2^2},~~~~dc_i=0,
\eeq
and as a result, every function has be solved in terms of $\rho$ and the four integration constants $c_i$.
We are now ready to calculate the fluxes. The non-trivial ones take the form
\begin{align}\label{eq:newIIB2}
B &= \sqrt{c_3}\frac{1-c_4}{4 c_4 c_2}\rho^2\text{Vol}(S^2)~,~~~
F_3=\frac{1}{4} \nu c_1^2 c_2 (1-c_4)d\psi\wedge\text{Vol}(S^2)+ 2 \nu c_1^2c_2 c_4\text{Vol}(S^3)~,\\[2mm]
F_5&=B\wedge F_3+c_1 c_2^2 \nu\sqrt{c_3}d\left(\frac{\rho^2}{2(c_3\rho^2-c_1^2 c_2^2 c_4)}(d\psi+\cos\theta d\phi)\right)+\frac{\nu c_1^2\sqrt{c_3}c_4}{2}d\big(\rho^2(d\psi+\cos\theta d\phi)\big)\wedge \text{Vol}(S^3)~, \nn
\end{align}
where $dB= H$. Clearly the Bianchi identites of the fluxes are implied automatically and one can show that this is true of \eqref{eq:IIBbpsfinalf} also. So this case contains a single example, expressed in terms of 4 integration constants. The 10d metric, warp factor and dilation then take the form
\begin{align}\label{eq:newIIB1}
ds^2&= e^{2A}ds^2(\mathbb{R}^{1,2})
+e^{-2A}\bigg[d\rho^2+\frac{1}{1-c_4}\rho^2 ds^2(S^3)+\frac{\rho^2}{4(c_4 - \frac{c_3}{c_1^2c_2^2}\rho^2)}(d\psi+\cos\theta d \phi)^2\nn\\[2mm]
&+ \frac{\rho^2 }{4 c_4 }ds^2(S^2)\bigg]~,~~~~
e^{-4A}= (1-c_4)\left(\frac{c_1c_4}{\rho^2}-\frac{c_3}{c_2^2}\right)~,~~~
e^{-4A+2\Phi}= \frac{1}{c_2^2}-\frac{c_3}{c_1^2c_2^4 c_4}\rho^2~.
\end{align}
This solution preserves an $SO(4)$ R-symmetry realised by one $SU(2)$ factor of the round $S^3$ and the $SU(2)$ of the squashed sphere - the residual symmetries of the spheres make up an $SU(2)\times U(1)$ flavour symmetry. Despite our assumption that $\alpha+\delta\neq \frac{\pi}{2}$ (which is when $c_3=0$) when deriving \eqref{eq:IIBbpsfinala}-\eqref{eq:IIBbpsfinalf} there is in fact no issue with taking this limit, which merely collapses this solution to that of section \ref{sec: additional S3}. There is good reason for this, as one can actually generate this solution from section \ref{sec: additional S3} by first T-dualising on $\partial_{\psi}$ then performing a formal U-duality\footnote{T-dualities along the spatial components of Mink$_3$, then a lift to M-theory followed by a boost along the M-theory $U(1)$ before reducing to IIA, and finally undoing the spatial T-dualities. This process needs to be supplemented by rescaling the coordinates along the way, which is why we refer to this as a formal U-duality.} on the Mink$_3$  followed by another T-duality on $\partial_{\psi}$. Additionally, this solution is also contained in section \ref{sec:novel}: it can be obtained by imposing that the coordinate $x$ there (which should be identified with  $\psi$ in this section) is an isometry and then T-dualising it.\\

This concludes our IIB classification, we shall now turn our attention towards IIA.

\section{Mink$_3$ with an $S^3$ factor in IIA}\label{IIA}
The type IIA supersymmetry conditions obtained from plugging
\eqref{eq:7d-bispinors} into the seven-dimensional supersymmetry constraints \eqref{7dsusy} lead to the following constraints on the four-dimensional bispinors
\begin{subequations}\label{eq: IIaSUSY}
\begin{align}
&d_{H_3}(e^{2A-\Phi}\text{Im}\psi^1_-)=0~,\label{eq: IIaSUSYa}\\[2mm]
&d_{H_3}(e^{2A+2C-\Phi}\psi^2_-)+ 2i\nu e^{2A+C-\Phi}\psi^2_{\hat\gamma+}=0~,\label{eq: IIASUSYb}\\[2mm]
&d_{H_3}(e^{3A+2C-\Phi}\psi^2_+)+ 2i\nu e^{3A+C-\Phi}\psi^2_{\hat\gamma-}=0~,\label{eq: IIASUSYc}\\[2mm]
&d_{H_3}(e^{2A+2C-\Phi}\text{Re}\psi^1_-)- 2\nu e^{2A+ C-\Phi}\text{Im}\psi^1_{\hat\gamma+}=0~,\label{eq: IIASUSYd}\\[2mm]
&d_{H_3}(e^{2A+3C-\Phi}\text{Re}\psi^1_{\hat\gamma_+})-e^{2A+3C-\Phi}H_0 \text{Im}\psi^1_-=0~,\label{eq: IIASUSYe}\\[2mm]
&d_{H_3}(e^{3A+2C-\Phi}\text{Im}\psi^1_+)+ 2\nu e^{3A+ C-\Phi}\text{Re}\psi^1_{\hat\gamma-}=0~,\label{eq: IIASUSYf}
\end{align}
\end{subequations}
while the fluxes are defined through
\begin{subequations}\label{eq: IIaSUSY_fluxes}
\begin{align}
&d_{H_3}(e^{3A-\Phi}\text{Re}\psi^1_+)+ e^{3A}\star_4 \lambda(G_-)=0~,\\[2mm]
&d_{H_3}(e^{3A+ 3C-\Phi}\text{Im}\psi^1_{\hat\gamma-})+  e^{3A+ 3C-\Phi}H_0 \text{Re}\psi^1_++\nu e^{3A+3C}\star_4 \lambda(G_+)=0~,
\end{align}
\end{subequations}
and must additionally satisfy
\beq
\bigg(\text{Re}\psi^1_{\hat\gamma+}\wedge\lambda(G_+)+ \text{Im}\psi^1_-\wedge \lambda(G_-)\bigg)\bigg \lvert_4=0
\eeq
As before, we will first examine the 0-form constraints. These are given by two of the three constraints that were found for type IIB:
\beq
(\psi^2_{\hat\gamma})_0 = (\text{Im}\psi^1_{\hat\gamma})_0=0 \;.
\eeq
Again, the solutions branch off similar to type IIB, with an $\a = 0$ and an $\a \neq 0$ branch.
We parameterise Branch I as in \eqref{eq: branchoneparameterisation} and Branch II as in \eqref{eq:branchIIparamaterisation}.
\subsection{Branch I: Solutions with $\alpha=0$}
As was the case in IIB  we first study the lower form conditions that follow from \eqref{eq: IIbSUSY}. After once more rotating the canonical frame of \eqref{eq: canonicalframe} by \eqref{eq: killframe} we extract the necessary, but not sufficient, supersymmetry  conditions
\begin{subequations}
\begin{align}
&d( e^{2A+3C-\Phi} \cos\beta \sin\delta)+  e^{2A+3C-\Phi} H_0\cos\beta w_1=0~,\label{eq:branchIsimpcondIIA1}\\[2mm]
&e^{2A+3C-\Phi} H_0\sin\beta (\cos\delta w_1-\sin\delta v_2)\wedge v_1\wedge w_2+ \cos\beta(...) =0~,\label{eq:branchIsimpcondIIA5}\\[2mm]
&d(e^{2A-\Phi}\cos\beta w_1)=d(e^{3A+2C-\Phi}\cos\beta \cos\delta)- 2\nu e^{3A+C-\Phi} \cos\beta v_2=0~,\label{eq:branchIsimpcondIIA2}\\[2mm]
&\cos\beta (\sin\delta v_1+ \frac{\nu}{2} e^{C}\cos\delta d\theta)=\cos\beta (\sin\delta w_2-\frac{\nu}{2}e^{C}\cos\delta \sin\theta d\phi) =0~,\label{eq:branchIsimpcondIIA3}\\[2mm]
&d(e^{2A+2C-\Phi} \sin\beta(\cos\delta w_1- \sin\delta v_2))+ 2\nu e^{2A+C-\Phi}(\sin\beta w_1\wedge v_2-\cos\beta \sin\delta v_1\wedge w_2)\nn\\[2mm]
&-\cos\beta\cos\delta e^{2A+2C-\Phi}(d\theta\wedge w_2+ \sin\theta d\phi \wedge v_1)=0,\label{eq:branchIsimpcondIIA4}
\end{align}
\end{subequations}
where $ \cos\beta(...) $ represents further terms which vanish when $\cos\beta=0$. These are sufficient to  truncate the ansatz considerably.  We note from \eqref{eq:branchIsimpcondIIA1} that if $\sin\delta=0$ then either $H_0=0$ or $\cos\beta=0$, however the latter also leads to $H_0=0$ because of \eqref{eq:branchIsimpcondIIA5} - so $\sin\delta=0$ implies $H_0=0$. We also observe that if $\sin\delta=0$ then \eqref{eq:branchIsimpcondIIA2} requires $d\theta=d\phi=0$, or naively $\cos\theta=0$ but this is a subcase of the former (see \eqref{eq: branchoneparameterisation}), so $\sin\delta=0$ also implies $d\theta=d\phi=0$. Our task now is to show that, as in IIB, $\sin\delta=0$ is a necessary condition: first we note that if we set $\cos\delta=0$ then there is no solution as \eqref{eq:branchIsimpcondIIA2} sets the vielbein to zero, thus we can restrict our considerations to $0<\sin\delta<\frac{\pi}{2}$ where $(w_1,~v_2)$ must span an $S^2$. However, as was the case in IIB, \eqref{eq:branchIsimpcondIIA4} can be rewritten as
\beq
d(e^{2A+2C-\Phi} \sin\beta(\cos\delta w_1- \sin\delta v_2))+ 2\nu e^{2A+C-\Phi}(\sin\beta w_1\wedge v_2+\cos\beta \sin\delta v_1\wedge w_2)=0,
\eeq
using \eqref{eq:branchIsimpcondIIA3} which excludes this because $w_1\wedge v_2$ gives rise to a $\text{Vol}(S^2)$ term that can not be cancelled by the parts involving $(v_1,w_2)$. Thus we can once more conclude that
\beq
H_0= \delta= \theta= \phi=0~.
\eeq
Given this, we can write the necessary and sufficient solutions for supersymmety in the $\alpha=0$ limit in a relatively simple way. After rotating the canonical frame, this time  as in \eqref{eq:branchIsimplerot},  we find
\begin{align}
&H_0=d(e^{2A-\Phi}\cos\beta v_1)= d(e^{\frac{3}{2}A+C-\frac{\Phi}{2}}\sqrt{\cos\beta})-\nu e^{\frac{3}{2}A-\frac{\Phi}{2}}\sqrt{\cos\beta}v_2= 0~,\nn \\[2mm]
&d(e^{2A+2C-\Phi} w)+ 2\nu e^{2A+C-\Phi} w\wedge v_2=d(e^{2A+2C-\Phi}\sin\beta v_1)+ 2\nu e^{2A+C-\Phi} \sin\beta v_1\wedge v_2=0~,\nn\\[2mm]
&d(e^{3A+2C-\Phi} v_1\wedge w)- 2\nu e^{3A+2C-\Phi} v_1\wedge w\wedge v_2=d(e^{3A-\Phi}\cos\beta)\wedge v_1\wedge v_2=0~,\nn\\[2mm]
&d(e^{3A+2C-\Phi}\sin\beta w_1\wedge w_2)- 2\nu e^{3A+2C-\Phi} \sin\beta w_1\wedge w_2\wedge v_2+ e^{3A+2C-\Phi}\cos\beta H_3=0~,\nn\\[2mm]
&e^{2A}H_3\wedge v_1+\cos^2\beta d(e^{2A}\tan\beta)\wedge v_1\wedge w_1\wedge w_2 =d(e^{-A+\Phi}\sin\beta)\wedge v_1 \wedge w_1\wedge w_2= w \wedge H_3=0~,\nn\\[2mm]
&d(e^{3A-\Phi}\sin\beta)+ d(e^{3A-\Phi}\cos\beta w_1\wedge w_2)-e^{3A-\Phi}\sin\beta H_3-e^{3A}\star_4 \lambda(G_-)=0~,\nn\\[2mm]
&d(e^{3A+3C-\Phi} \sin\beta v_2)+ d(e^{3A+3C-\Phi}\cos\beta w_1\wedge w_2\wedge v_2)-e^{3A+3C-\Phi}\sin\beta H_3\wedge v_2-e^{3A+3C}\star_4 \lambda(G_+)=0~,\nn\\[2mm]
&\bigg(\cos\beta v_1\wedge v_2\wedge \lambda(G_+)+(\cos\beta-\sin\beta w_1\wedge w_2)\wedge v_1\wedge \lambda(G_-) \bigg)\bigg\lvert_{4}=0~.\label{eq: generalbranchIsusyIIA}
\end{align}
This is as far as we can go without making assumptions about $\beta$, which we now proceed to do.

\subsubsection{Sub Case: $\beta=0$}\label{sec:branchIbeta0}
Setting $\beta=0$ in  \eqref{eq: generalbranchIsusyIIA} immediately leads to $H_3=0$, the rest of the conditions are implied by
\begin{subequations}
	\begin{align}
	& d(e^{\frac{3}{2}A+C-\frac{\Phi}{2}})-\nu e^{\frac{3}{2}A-\frac{\Phi}{2}}v_2 = d(e^{2A-\Phi}v_1)=d(e^{-A}w) = 0~, \label{eq:IIA_al0b10a10_a}\\[2mm]
	& d(e^{2A})\wedge v_1\wedge v_2 = d(e^{-A+\Phi})\wedge w\wedge v_1 =0~,\label{eq:IIA_al0b10a10_b}\\[2mm]
	& e^{3A}\star_4 \lambda(G_-)- d(e^{3A-\Phi} w_1\wedge w_2) = 0~, \label{eq:IIA_al0b10a10_c}\\[2mm]
	& e^{3A+3C}\star_4\lambda(G_+)- d( e^{3A+3C-\Phi} w_1\wedge w_2\wedge v_2 )=0~, \label{eq:IIA_al0b10a10_d}\\[2mm]
	& \bigg( v_1\wedge v_2\wedge \lambda(G_+)+ v_1\wedge \lambda(G_-) \bigg)\bigg\lvert_{4}=0~. \label{eq:IIA_al0b10a10_e}
	\end{align}
\end{subequations}
The first thing we note is that given \eqref{eq:IIA_al0b10a10_c}-\eqref{eq:IIA_al0b10a10_d}, $G_+$ must be a 0-form and $G_-$ a 1-form which means \eqref{eq:IIA_al0b10a10_e} is solved automatically. Next, we solve \eqref{eq:IIA_al0b10a10_a} by using it to define the vielbein
\begin{equation}
v_1= e^{-2A+\Phi}g(x) dx,~~~ w= e^A( d\psi_1+ i d\psi_2)~,~~~v_2 = \nu\, e^{-3A/2+\Phi/2}d\rho,~~~ \rho= e^{\frac{3}{2}A+C-\frac{\Phi}{2}}~,
\end{equation}
where $g$ is a function parametrising a potential coordinate transformation in $x$.
From eq.~\eqref{eq:IIA_al0b10a10_c}, we see that the combination $e^{A-\Phi}$  only depends on $x$
and that  $e^A$, $e^\Phi$ and $e^C$ are functions of  $x, \rho$ only, so that $\partial_{\psi_1}$ and $\partial_{\psi_2}$ are isometries. We thus choose to parametrise
\beq\label{d4d83}
e^{A-\Phi}= f(x)~,~~~g= -f~.
\eeq
the latter of which is a convenient choice we make without loss of generality.
For the fluxes, we use the $v_1, v_2, w_1, w_2$ vielbein on $M_4$ to compute the Hodge duals from eq.~\eqref{eq:IIA_al0b10a10_c} and \eqref{eq:IIA_al0b10a10_d},  arriving at the ten-dimensional fluxes
\beq\label{eq:branch1IIAbeta0Fluxes}
F_0 = \partial_x f~,~~~~~F_4 = \nu \rho^3\bigg(f\partial_{\rho}(f^{-1}e^{-4A}) dx-  \partial_x(f^{-1}e^{-4A})d\rho\bigg)\wedge\text{Vol}(S^3)~,
\eeq
The  Bianchi identities reduce to $dF_0=dF_4=0$ which leads to
\beq\label{eq:branch1IIAbeta0PDEs}
\partial_{x}^2 f= 0~,~~~\partial_x^2(f^{-1}e^{-4A})+f \frac{1}{\rho^3}\partial_{\rho}(\rho^3 \partial_{\rho}(f^{-1}e^{-4A}))=0~,
\eeq
the former of which can be immediately integrated as $f =(c+ F_0 x)$, $dc=0$. The metric takes the form
\beq\label{d4d81}
ds^2= \frac{1}{\sqrt{f H}} ds^2(\mathbb{R}^{1,4}) +\sqrt{\frac{H}{f}}\bigg(d\rho^2+ \rho^2 ds^2(S^3)\bigg)+\sqrt{f H} dx^2,~~~ H= f^{-1} e^{-4A}.
\eeq
This solution corresponds to an intersecting D4-D8 brane system, where the localised D4-branes are embedded in the D8-branes \cite{youm}.

\subsubsection{Sub Case: $\beta=\frac{\pi}{2}$}\label{sec:IIA_a0_bpi2}
The $\beta=\frac{\pi}{2}$ limit of \eqref{eq: generalbranchIsusyIIA} leads to $H_3 \propto v_1\wedge w_1\wedge w_2$ with the remaining conditions implied by
\begin{subequations}
	\begin{align}
	&  d( e^{2A+2C-\Phi}u_i)+ 2\nu e^{2A+C-\Phi}u_i\wedge v_2= 0~, \label{eq:IIA_al0b10a11_a}\\[2mm]
	&  d( e^{3A+2C-\Phi}\epsilon_{ijk}u_j\wedge u_k)- 2\nu e^{3A+C-\Phi}\epsilon_{ijk}u_j\wedge u_k\wedge v_2= 0~, \label{eq:IIA_al0b10a11_b}\\[2mm]
	& d(e^{3A-\Phi})-e^{3A-\Phi} H_3+ e^{3A}\star_4 \lambda(G_-)=0~, \label{eq:IIA_al0b10a11_d}\\[2mm]
	&d(e^{3A+3C-\Phi} v_2)-e^{3A+3C-\Phi} H_3\wedge v_2+ e^{3A+3C}\star_4 \lambda(G_+)=0~, \label{eq:IIA_al0b10a11_e},\\[2mm]
  &d(e^{-A-\Phi})\wedge v_1\wedge w_1\wedge w_2=v_1\wedge w_1\wedge w_2\wedge \lambda(G_-)  \bigg\lvert_{4} =0~,  \label{eq:IIA_al0b10a11_f}
	\end{align}
\end{subequations}
where we introduce
\beq
u=(v_1,w_1,w_2).
\eeq
to ease notation, and to make clear the cyclic property of these vielbein.
The first thing we note is that, given \eqref{eq:IIA_al0b10a11_d}, the second of \eqref{eq:IIA_al0b10a11_f} reads $\star_4 H_3\wedge v_1\wedge w_1\wedge w_2=0$, but since $H_3\propto  v_1\wedge w_1\wedge w_2$ we must set
\beq
H_3=0~.
\eeq
Next one can show that both  \eqref{eq:IIA_al0b10a11_a} and \eqref{eq:IIA_al0b10a11_b} together imply the useful identities
\beq
d(e^{A}u_i)\wedge u_j =\big[d(e^{A/2+C-\Phi/2})-\nu e^{A/2-\Phi/2} v_2\big]\wedge u_i\wedge u_j=0~,~~~ i\neq j,
\eeq
so we must have
\beq\label{eq:IIabranchIbetapiovertowveil}
d(e^{A}u_i ) +\frac{c_i}{2} \epsilon_{ijk}u_j\wedge u_k=d(e^{A/2+C-\Phi/2})- \nu e^{A/2-\Phi/2} v_2=dc_i=0~.
\eeq
Consistency of the first of these with \eqref{eq:IIA_al0b10a11_a} implies that $c_i=0$, and with this fixed \eqref{eq:IIA_al0b10a11_b} is also follow from \eqref{eq:IIabranchIbetapiovertowveil}. We can use the standard trick of taking \eqref{eq:IIabranchIbetapiovertowveil} to define a vielbein in terms of local coordinates, namely
\beq\label{eq:IIabranchIbetapiovertowveil2}
u_i = e^{-A}dx_i,~~~v_2= \nu e^{-A/2+\Phi/2} d\rho,~~~~ \rho = e^{A/2+C-\Phi/2}.
\eeq
Having defined the vielbein, it is then a simple matter to solve the first of \eqref{eq:IIA_al0b10a11_f} by introducing a free function
\beq \label{d2d64}
e^{-A-\Phi}= f(x_1,x_2,x_3).
\eeq
All that remains is to calculate the fluxes, and impose their Bianchi identities. Using \eqref{eq:IIabranchIbetapiovertowveil2} to take the Hodge duals of \ref{eq:IIA_al0b10a11_d} and \ref{eq:IIA_al0b10a11_e} we find the 10d fluxes
\begin{align}\label{d2d62}
F_2&=\frac{1}{2}\epsilon_{ijk}\partial_{x_i}f dx^j\wedge dx^k,\\[2mm]
F_6&=-\nu \rho^3\bigg(\frac{1}{2}\epsilon_{ijk}\partial_{x_i}(f^{-1}e^{-4A})dx^j\wedge dx^k\wedge d\r-f\partial_{\rho}(f^{-1}e^{-4A})dx_1\wedge dx_2\wedge dx_3\bigg)\wedge \text{Vol}(S^3),\nn
\end{align}
which clearly means the Bianchi identities, away from localised sources, follow from
\begin{align}\label{d2d63}
\partial_{x_i}^2 f=0~,~~~~ \partial_{x_i}^2(f^{-1} e^{-4A})+ f \frac{1}{\rho^3}\partial_{\rho}(\rho^3\partial_{\rho}(f^{-1} e^{-4A})))=0~.
\end{align}
The metric takes the form
\eq{\label{d2d61}
ds^2&= \frac{1}{\sqrt{f H}} ds^2(\mathbb{R}^{1,2}) +\sqrt{\frac{H}{f}}\left(d\rho^2+ \rho^2 ds^2(S^3)\right)+\sqrt{f H} (dx_1^2+dx_2^2+dx_3^3)~,\\
   H&= f^{-1} e^{-4A}~.
}
The solution corresponds to an intersecting D2-D6 brane system \cite{ity}.\\

\subsubsection{Sub Case: Generic $\beta$}
For $0<\beta<\frac{\pi}{2}$ one is able to divide by $\sin\beta, \cos\beta$ freely when simplifying \eqref{eq: generalbranchIsusyIIA}. Assuming that $\cos\beta\neq 0$ the result is
\begin{subequations}
	\begin{align}
&d(e^{2A-\Phi}\cos\beta v_1)=  d(e^{-A}\sec\beta w )=0~,\label{eq:IIAbranchIgen1}\\[2mm]
&d(e^{\frac{3}{2}A+C-\frac{1}{2}\Phi} \sqrt{\cos\beta} )-\nu  e^{\frac{3}{2}A-\frac{1}{2}\Phi} \sqrt{\cos\beta}v_2=0~,\label{eq:IIAbranchIgen2}\\[2mm]
&d(e^{-A+\Phi}\sec\beta)\wedge v_1\wedge w=d(e^{2A})\wedge v_1\wedge v_2= d(e^{-2A}\tan\beta)\wedge v_1=0~,\label{eq:IIAbranchIgen3}\\[2mm]
&e^{2A}\cos^2\beta H_3+ d(e^{2A}\sin\beta \cos\beta)\wedge w_1\wedge w_2=0~,\label{eq:IIAbranchIgen4}\\[2mm]
&d(e^{3A+3C-\Phi} \sin\beta v_2)+ d(e^{3A+3C-\Phi}\cos\beta w_1\wedge w_2\wedge v_2)-e^{3A+3C-\Phi}\sin\beta H_3\wedge v_2-e^{3A+3C}\star_4 \lambda(G_+)=0\nn,\\[2mm]
&d(e^{3A-\Phi}\sin\beta)+ d(e^{3A-\Phi}\cos\beta w_1\wedge w_2)-e^{3A-\Phi}\sin\beta H_3-e^{3A}\star_4 \lambda(G_-)=0~,\label{eq:IIAbranchIgen5}\\[2mm]
&\bigg(\cos\beta v_1\wedge v_2\wedge \lambda(G_+)+(\cos\beta-\sin\beta w_1\wedge w_2)\wedge v_1\wedge \lambda(G_-) \bigg)\bigg\lvert_{4}=0~,\label{eq:IIAbranchIgen7}
\end{align}
\end{subequations}
where $H_3$ is closed given \eqref{eq:IIAbranchIgen1} and \eqref{eq:IIAbranchIgen3}.
As usual we solve \eqref{eq:IIAbranchIgen1} and \eqref{eq:IIAbranchIgen2} by using them to define a vielbein in terms of local coordinates
\begin{align}
v_1 &=g(x)e^{-2A+\Phi}\sec\beta dx,~~~  w = e^{A} \cos\beta (d\psi_1+i d\psi_2),\nn\\[2mm]
v_2&=\nu  e^{-\frac{
3}{2}A+\frac{1}{2}\Phi} \sqrt{\sec\beta}d\rho,~~~ \rho=e^{\frac{3}{2}A+C-\frac{1}{2}\Phi} \sqrt{\cos\beta}.
\end{align}
where $g(x_1)$ is a function parametrising a potential diffeomorphism in $x_1$.
With local coordinates introduced we can solve \eqref{eq:IIAbranchIgen3} in terms of them as
\beq\label{d4u3}
e^{A-\Phi}\cos\beta= f(x)~,~~~ A= A(\rho,x)~,~~~\tan\beta= c(x_1)e^{2A}~,~~~ g=-f~,
\eeq
so that $\partial_{\psi_i}$ are necessarily isometry directions.
We can then calculate the ten-dimensional fluxes as before - first we note that
\beq\label{d4u2a}
F_0= \partial_{x}f+ \frac{\partial_{x}c f\tan^2\beta}{c}~.
\eeq
should be constant. We shall restrict ourselves to the case $\p_x c = 0$.  For generic $\beta$ then the fluxes may be expressed as
\begin{align}\label{d4u2b}
F_0&=\partial_{x}f,~~~F_2= F_0 B_2,~~~ B_2=-\frac{\sin^2\beta}{c}d\psi_1\wedge d\psi_2,~~~ dc=0\\[2mm]
F_4&=B_2\wedge F_2+ \nu \rho^3\bigg(f\partial_{\rho}(f^{-1}e^{-4A}) dx-  \partial_x(f^{-1}e^{-4A})d\rho\bigg)\wedge\text{Vol}(S^3)\nn.
\end{align}
We note that the Bianchi identities follow when anything not coupled to $B_2$ is closed, and since these terms reproduce \eqref{eq:branch1IIAbeta0Fluxes} the Bianchi identities imply the PDEs of \eqref{eq:branch1IIAbeta0PDEs} once more. This is because the class of solutions in this section can be generated via U-duality, from the intersecting D4-D8 system in section \ref{sec:branchIbeta0}. For completeness the metric takes the form
\beq\label{d4u1}
ds^2= \frac{1}{\sqrt{f H}} ds^2(\mathbb{R}^{1,2})+\frac{\cos^2\beta}{\sqrt{f H}} ds^2(T^2) +\sqrt{\frac{H}{f}}\bigg(d\rho^2+ \rho^2 ds^2(S^3)\bigg)+\sqrt{f H} dx^2,~~~ H= f^{-1} e^{-4A},
\eeq
where $T^2$ is spanned by $(\psi_1,~\psi_2)$.

\subsection{Branch II: $\alpha$ non-Zero}
For the second branch with $0<\alpha<\pi$, we begin by studying the lower form conditions that follow from \eqref{eq: IIaSUSY}. As in IIB we find it useful to rotate the canonical frame of \eqref{eq: canonicalframe} as  \eqref{eq: branchIIrot}. Necessary but insufficient conditions for supersymmetry are
\begin{subequations}
\begin{align}
&d(e^{2A+3C-\Phi}\cos\beta\cos(\alpha-\delta))+e^{2A+3C-\Phi}H_0(\cos\beta \cos\delta v_2+ \sin\beta w_2)=0,\label{eq:IIABIIa}\\[2mm]
&d(e^{2A-\Phi}(\cos\beta\cos\delta v_2+ \sin\beta w_2))=0,\label{eq:IIABIIb}\\[2mm]
&d(e^{3A+2C-\Phi}\cos\alpha \sin\beta)+ 2 \nu e^{3A+C-\Phi}(\cos\beta \cos\delta w_2- \sin\beta v_2)=0,\label{eq:IIABIIc}\\[2mm]
&e^{C}\cos\alpha\sin\beta \sin\theta d\phi-2\nu(\cos\beta \cos\delta w_1- \sin\alpha \sin\beta v_1)=0,\label{eq:IIABIId}\\[2mm]
&e^{C}\cos\alpha\sin\beta d\theta-2\nu(\cos\beta \cos\delta v_1+ \sin\alpha \sin\beta w_1)=0,\label{eq:IIABIIe}\\[2mm]
&d(e^{2A+2C-\Phi}(\sin\alpha\sin\beta v_2-\sin(\alpha-\delta)\cos\beta w_2))\nn\\[2mm]
&+ 2 e^{2A+C-\Phi}\nu\cos\beta (\sin\delta w_2\wedge v_2+ \cos(\alpha-\delta) v_1\wedge  w_1)=0,\label{eq:IIABIIf}
\end{align}
\end{subequations}
First from \eqref{eq:IIABIIa} we observe  that when $\cos(\alpha-\delta)=0$ ($\cos\beta=0$ is a subcase of this) we necessary have $H_0=0$. Next we observe that generically \eqref{eq:IIABIIb}-\eqref{eq:IIABIIe} can be used to locally define the vielbein on $M_4$, the only exception is when $\sin\beta=0$  ($\cos\alpha=0$ is a subcase of this).

Setting $\sin\beta=0$ means that in order to solve \eqref{eq:IIABIId}-\eqref{eq:IIABIIe} we must take $\delta=\frac{\pi}{2}$, additionally $(\theta,\phi)$ drop out of the definition of the spinors so we can fix
\beq
\beta=\theta=\phi=(\delta-\frac{\pi}{2})=0
\eeq
 without loss of generality. Interestingly one doesn't need to set $H_0=0$, however as we shall see in section \ref{sec:branchIIbetazroIIA} this case actually contains no solutions.

If one assumes $\sin\beta\neq0$ we see that $v_1,~w_1$ must span $S^2$ while $v_2, w_2$ can be expressed in terms of local coordinates in such a way that they have no legs in $S^2$. This is a problem for \eqref{eq:IIABIIf} which generically has an $\text{Vol}(S^2)$ factor, due to the $v_1\wedge w_1$ term which sits orthogonal to everything else. Thus the only resolution is to fix $\cos(\alpha-\delta)=0$ which leads to $H_0=0$ also. This actually leads to a novel class of solutions that we shall derive in section \ref{sec:novel}.

\subsubsection{Sub Case: $\beta=0$}\label{sec:branchIIbetazroIIA}
Upon setting $\beta=0$ we are lead to the following conditions for supersymmetry
\begin{subequations}
	\begin{align}
	&d(e^{2A+3 C-\Phi}\sin\alpha)= d(e^{-A-\Phi}\cos^4\alpha)\wedge v_1\wedge w_1 \wedge w_2 =0~,\label{eq:secondlasta}\\[2mm]
	&d(e^{2A+2 C-\Phi}\cos\alpha u_i)+\nu e^{2A+C-\Phi}(\sin\alpha\epsilon_{ijk}u_j\wedge u_k +2 u_i\wedge v_2)=0~,\label{eq:secondlastb}\\[2mm]
		& d(e^{3A+2 C-\Phi}(\epsilon_{ijk} u_j\wedge u_k +2 \sin\alpha u_i \wedge v_2))- 2\nu e^{3A+ C-\Phi}\cos\alpha \epsilon_{ijk} u_j\wedge u_k \wedge v_2=0~,\label{eq:secondlastc}\\[2mm]
	&	\sin\alpha H_3- \cos\alpha H_0 v_1\wedge w_1\wedge w_2 =0~,\label{eq:secondlastd}\\[2mm]
	&e^{3A}\star_4 \lambda (G_-)+ d(e^{3A-\Phi})- e^{3A-\Phi}H_0 \cot\alpha v_1\wedge w_1\wedge w_2=0~,\label{eq:secondlaste}\\[2mm]
		&e^{3A+3C}\star_4 \lambda (G_+)+ e^{3A+3C-\Phi}H_0+ d(e^{3A+3C-\Phi}\cos\alpha v_2)- e^{3A+3C-\Phi}H_0 \csc\alpha v_1\wedge v_2\wedge w_1\wedge w_2 =0 \nn,\\[2mm]
&	\bigg(\cos\alpha  v_1\wedge w_1\wedge w_2\wedge \lambda(G_-)- (\sin\alpha- v_1\wedge v_2\wedge w_1\wedge w_2)\wedge \lambda(G_+)\bigg)\bigg\lvert_{4}=0~\label{eq:secondlastf}.
		\end{align}
	\end{subequations}
	where here as elsewhere $u=(v_1,w_1,w_2)$. Using the same techniques as are spelled out in section \ref{eq:D5case}, it is possible to establish that
\beq
d(e^{\frac{1}{2}A+C-\frac{1}{2}\Phi})-\nu e^{\frac{1}{2}A-\frac{1}{2}\Phi}\cos\alpha v_2 = d(e^{A} u_i)\wedge u_j =0, ~~\text{for}~i\neq j
\eeq
which we can use as in section \ref{sec: additional S3} to define the vielbein in terms of the local coordinate $\rho= e^{\frac{1}{2}A+C-\frac{1}{2}\Phi}$ and a set of left invarient 1-forms such that
\beq
v_2 = \nu e^{-\frac{1}{2}A+\frac{1}{2}\Phi}d\rho,~~~ u_i = e^{-A}\cos\alpha c_i\tilde{K}_i,~~~d\tilde{K}_i=\frac{1}{2}\tilde{K}_j\wedge \tilde{K}_k,
\eeq
under the assumption that $\alpha\neq 0$.
Plugging this back into \eqref{eq:secondlastb} fixes
\beq
e^{2A} = \frac{c^4e^{-2\Phi} \sin^4\alpha}{\rho^4},~~~ \sin\alpha =\frac{c_1}{\rho},~~c_i=c
\eeq
however plugging this back into \eqref{eq:secondlasta} leads to
\beq
d\rho\wedge \tilde{K}_1\wedge \tilde{K}_2\wedge \tilde{K}_3=0
\eeq
which cannot be solved.

\subsubsection{Sub Case: $\delta=\alpha+\frac{\pi}{2}$ }\label{sec:novel}
The final case we consider is when $\delta=\alpha+\frac{\pi}{2}$ and contains $\beta=\frac{\pi}{2}$ as a sub case. In addition to the rotating the canonical frame \eqref{eq: canonicalframe} by \eqref{eq: branchIIrot} we find it useful to send $v_1+ i w_1 \to e^{-i\beta}(v_1+ i w_1)$, then the necessary and sufficient conditions for supersymmetry are
\begin{subequations}
\begin{align}
&d(e^{A-C}\sin\alpha)=H_0=0~,\label{eq:BPSfinal1}\\[2mm]
&d(e^{3A+2C-\Phi}\cos\alpha\sin\beta)- 2\nu e^{3A+C-\Phi}k_1=d(e^{2A-\Phi}k_2)=0~.,\label{eq:BPSfinal2}\\[2mm]
&e^{C}\cos\alpha\sin\beta d\theta+ 2\nu\sin\alpha v_1=e^{C}\cos\alpha\sin\beta \sin\theta d\phi+ 2\nu\sin\alpha w_1=0~,\label{eq:BPSfinal2}\\[2mm]
&H_3=\frac{1}{4}\left(d(e^{2C}\cot^2\alpha\cos\beta \sin\beta)-2 e^{C}\nu \cos\alpha w_2\right)\wedge \text{Vol}(S^2),\label{eq:BPSfinal3}\\[2mm]
&d\left(\frac{e^{-A+C}\sin\alpha}{\Delta}\right)\wedge k_1-e^{2A+C-\Phi}\cos\alpha \sin\beta d\left(\frac{e^{-3A+\Phi}\cos\alpha\cos\beta}{\Delta}\right)\wedge k_2=0~,\label{eq:BPSfinal4}\\[2mm]
&d\left(\frac{e^{-\frac{3}{2}A+C+\frac{1}{2}\Phi}\cos^{\frac{3}{2}}\alpha\sqrt{\sin\beta}\cos\beta}{\Delta}\right)\wedge k_1-e^{\frac{1}{2}A-\frac{1}{2}\Phi}\sqrt{\cos\alpha \sin\beta} d\left(\frac{e^{-2A+C+\Phi}\cot\alpha\sin\beta}{\Delta}\right)\wedge k_2=0~,\nn\\[2mm]
&e^{3A}\star_4 \lambda(G_-)+d(e^{3A-\Phi}\cos\alpha\cos\beta)-d(e^{3A-\Phi}\cos\alpha\sin\beta v_1\wedge w_1)-e^{3A-\Phi}\cos\alpha\cos\beta H_3=0~,\nn\\[2mm]
&e^{3A+3C}\star_4 \lambda(G_+)+d(e^{3A+3C-\Phi}k_3)-d(e^{3A+3C-\Phi}v_1\wedge w_1\wedge k_1)+ e^{3A+3C-\Phi}k_2\wedge H_3,\label{eq:BPSfinal5}\\[2mm]
&\bigg(\cos\alpha\sin\beta(1\!+\!\cot\beta v_1\wedge w_1)\wedge v_2\wedge w_2 \wedge\lambda(G_+)\!-\!(k_2\!-\! k_4\wedge v_1 \wedge w_1)\wedge\lambda(G_-)\bigg)\Big\lvert_{4}=0\label{eq:BPSfinal6}.
\end{align}
\end{subequations}
where we introduce
\begin{align}
k_1&=(\sin\alpha\cos\beta w_2+ \sin\beta v_2),~~~k_2=(\sin\alpha\cos\beta v_2-\sin\beta w_2),\nn\\[2mm]
k_3&=(\cos\alpha v_2-\sin\alpha\sin\beta w_2),~~~ k_4=(\cos\alpha w_2+\sin\alpha\sin\beta w_1),\nn\\[2mm]
\Delta&= \sin^2\beta+\cos^2\beta\sin^2\alpha,~~~ \text{Vol}(S^2)=\sin\theta d\theta\wedge d\phi,
\end{align}
to ease presentation. We can use \eqref{eq:BPSfinal2}-\eqref{eq:BPSfinal3} to locally define the vielbein through
\begin{align}\label{eq:lastveil}
k_1 &= \nu\sqrt{\cos\alpha \sin\beta}e^{-\frac{3}{2}A+\frac{1}{2}\Phi}d\rho,~~~k_2= e^{-2A+\Phi}dx,\\[2mm]
v_1&=-\frac{1}{2\sin\alpha}\nu e^{-\frac{3}{2}A+\frac{1}{2}\Phi}\rho\sqrt{\cos\alpha \sin\beta}d\theta,~~~w_1=-\frac{1}{2\sin\alpha}\nu e^{-\frac{3}{2}A+\frac{1}{2}\Phi}\rho\sqrt{\cos\alpha \sin\beta}\sin\theta d\phi,\nn
\end{align}
where $(\theta,\phi,x)$ and
\beq
\rho= e^{\frac{3}{2}A+C-\frac{1}{2}\Phi}\sqrt{\cos\alpha \sin\beta}
\eeq
are local coordinates on $M_4$. The ten-dimensional metric then takes the form
\begin{align}\label{nw1}
&ds^2 = \\
& e^{2A}ds^2(\mathbb{R}_{1,2}) + \frac{e^{-3A+\Phi}}{\cos\alpha \sin\beta}\rho^2ds^2(S^3) + \frac{e^{-4A+2\Phi}dx^2}{\Delta}+\frac{e^{-3A+\Phi}\cos\alpha \sin\beta}{4 \sin^2\alpha}\bigg(\frac{4 \sin^2\alpha d\rho^2}{\Delta}+\rho^2 ds^2(S^2)\bigg). \nn
\end{align}
We can now turn our attention to \eqref{eq:BPSfinal1} and  \eqref{eq:BPSfinal4} which lead to the PDEs
\begin{subequations}\label{nw3}
\begin{align}
&\partial_{\rho}(e^{A-C}\sin\alpha)=\partial_{x}(e^{A-C}\sin\alpha)=0~,\label{eq:lastPDE1}\\[2mm]
& \nu \rho\partial_{\rho}\left(\frac{e^{-3A+\Phi}\cos\alpha\cos\beta}{\Delta}\right)+\partial_{x}\left(\frac{e^{-A+C}\sin\alpha}{\Delta}\right)=0~,\label{eq:lastPDE2}\\[2mm]
& \nu\partial_{\rho}\left(\frac{e^{-2A+C+\Phi}\cot\alpha\sin\beta}{\Delta}\right)+ \partial_{x}\left(\frac{e^{-\frac{3}{2}A+C+\frac{1}{2}\Phi}\cos^{\frac{3}{2}}\alpha\sqrt{\sin\beta}\cos\beta}{\Delta}\right)=0~,\label{eq:lastPDE3}
\end{align}
\end{subequations}
and tell us that $e^{2A},~e^{2C},~e^{\Phi},~\alpha,~\beta$ are functions of $(\rho,~x)$ only, which means these solutions support an additional $SU(2)$ isometry due to round $S^2$ spanned by $(\theta,~\phi)$. Actually this $SU(2)$ is an additional part of an enhanced R-symmetry which together with the $SU(2)_R$ of $S^3$ gives $SO(4)_R$ - there is also an $SU(2)$ flavour symmetry. It is now a simple matter to calculate the Hodge dual of the fluxes from \label{eq:BPSfinal5} and then the fluxes themselves given \eqref{eq:BPSfinal3} and our local vielbein \eqref{eq:lastveil}. At first $G_{\pm}$ take a complicated form that we will not quote here, however, we are yet to impose also \eqref{eq:BPSfinal6}: doing so and making  extensive use  of \eqref{eq:lastPDE1}-\eqref{eq:lastPDE3} one can express the ten-dimensional fluxes as:
\begin{align}\label{eq:lastfluxes}
H &=\frac{1}{4}\left(d(e^{2C}\cot^2\alpha\cos\beta \sin\beta)-\frac{2}{\Delta}(\rho e^{-3A+\Phi}\cos\alpha\cos\beta d\rho- \nu e^{-2A+C+\Phi}\cot\alpha \sin\beta dx) \right)\wedge \text{Vol}(S^2)~,\nn\\[2mm]
F_0 &= 2 \frac{\Delta}{\sin\alpha}e^{-2\Phi} \partial_{x}(e^{2A}\sin\alpha)~,\nn\\[2mm]
F_2 &=\frac{e^{-3A-\frac{1}{2}\Phi}\rho \sqrt{\cos\alpha}}{4 \sin^2\alpha}\left(F_0 e^{\frac{3}{2}\Phi}\rho \sqrt{\cos\alpha} \cos\beta-2\nu e^{\frac{3}{2}A}\sin\alpha \sqrt{\sin\beta}\right)\text{Vol}(S^2)~,\nn\\[2mm]
F_4 &=
-\text{Vol}_3\wedge d(e^{3A-\Phi}\cos\alpha \sin\beta)
+\frac{e^{-\frac{5}{2}A+\frac{1}{2}\Phi}\rho^2}{\sqrt{\cos\alpha \sin\beta}\sin\alpha}
\bigg(d(e^{-2A}\cot\beta)\nn\\[2mm]
&-\frac{2e^{-\frac{5}{2}A}\sin\alpha}{\sin\beta \Delta}\big(\nu e^{\frac{1}{2}\Phi}\sqrt{\frac{\sin\beta}{\cos\alpha}}d\rho-e^{\frac{1}{2}A}\sin\alpha \cos\beta dx\big)\bigg)\wedge\text{Vol}(S^3)~,\nn\\[2mm]
F_6 &= - \text{Vol}_3\wedge (d(e^{3A-\Phi}\cos\alpha\sin\beta v_1\wedge w_1)+e^{3A-\Phi}\cos\alpha\cos\beta H_3)\\[2mm]
&+\frac{\nu e^{-10A+2\Phi}\rho^5}{4 \sin^2\alpha}\bigg(d(e^{3A-\Phi} \cos\alpha\cos\beta)\wedge d\rho- \frac{e^{-A+\Phi}}{\cos\alpha\sin\beta}d(e^{3A-\Phi} \cos\alpha\cos\beta)\wedge dx\bigg)\wedge \text{Vol}(S^2)\wedge\text{Vol}(S^3)\nn
\end{align}
where \eqref{eq:BPSfinal6} can be expressed in terms of $F_0$ as
\beq
\rho\sqrt{\sin\beta}\partial_{\rho}(\rho e^{2A}\sin\alpha)+\frac{\nu}{2}F_0\frac{e^{\frac{1}{2}A+\frac{3}{2}\Phi}\sin^2\alpha\sqrt{\cos\alpha} \cos\beta }{\Delta}=0\label{eq:lastparing}.
\eeq
Imposing that $F_0$ is constant together with \eqref{eq:lastPDE1}-\eqref{eq:lastPDE3} and \eqref{eq:lastparing} then implies $dH=0$ and the Biachi identities of the remaining fluxes. This system is quite complicated, however taking inspiration from section 4.1 of \cite{Macpherson:2016xwk} (which re-derives \cite{Imamura:2001cr}) we anticipate that the system can be further simplified if we treat the cases $F_0=0$ and $F_0\neq 0$ separately.

\subsubsection*{Subcase $F_0=0$}
If we set $F_0=0$ then \eqref{eq:lastparing} and \eqref{eq:lastfluxes} impose that
\beq\label{F0const1}
\rho\, e^{2A}\sin\alpha= L^2,~~~ dL=0
\eeq
This leaves \eqref{eq:lastPDE1}-\eqref{eq:lastPDE3} to solve. We first integrate \eqref{eq:lastPDE1} as
\beq\label{F0const2}
e^{A-C}\sin\alpha=c,~~~ dc=0,
\eeq
then use it to express \eqref{eq:lastPDE2} as
\beq
c \partial_x\left(\frac{e^{-5A+\Phi}}{\cos\alpha \sin\beta \Delta}\right)+ \nu \frac{1}{\rho}\partial_{\rho}\left(\frac{e^{-3A+\Phi} \cos\alpha \cos\beta}{\Delta}\right)=0.
\eeq
We note that this defines an integrability condition that we can solve by introducing an auxiliary function $h(\rho,x)$ such that
\beq
c\frac{e^{-5A+\Phi}}{\cos\alpha \sin\beta \Delta} =\frac{\nu}{\rho}\partial_{\rho}h,~~~ \frac{e^{-3A+\Phi} \cos\alpha \cos\beta}{\Delta}=-\partial_{x}h.
\eeq
Plugging these definitions into \eqref{eq:lastPDE3} and making use of \eqref{F0const1}-\eqref{F0const2} we arrive at
\beq
c^2\partial_{x}^2h + \frac{1}{\rho}\partial_{\rho}(\rho \partial_{\rho} h)=0.
\eeq
This is a 3d Laplacian expressed in axially symmetric cylindrical polar coordinates (up to rescaling $x$) . Solution in this class are in one to one correspondence with solution to this Laplace equation. The physical data can be expressed in terms of $h$ and the 2 constants $(c,~L)$, as
\begin{align}
L^4e^{-4A}&= \r^2\sin^2\alpha,~~~ e^{-\Phi+5A}=\frac{\nu\,c\,\rho}{\cos\alpha \sin\beta \partial_{\rho}h \Delta},~~\Delta= 1- \frac{c^3 \rho(\partial_{\rho}h)^2}{L^4 \partial_{\rho}h(1- c \rho \partial_{\rho}h)},\nn\\[2mm]
\tan\alpha&=\sqrt{\frac{L^4(1- c\rho \partial_{\rho}h)}{c^4 \rho^2 (\partial_{x}h)^2+ L^4(1-c\rho\partial_{\rho} h)^2}-1},\nn\\[2mm]
 \tan\beta &=\frac{\nu\sqrt{1-c\rho\partial_{\rho}h}\sqrt{L^4 \partial_{\rho}h (1- c\rho \partial_{\rho}h)- c^3 \rho (\partial_{x}h)^2)}}{c^{\frac{3}{2}}\sqrt{\rho}\partial_x h},
\end{align}
It is interesting that this class depends on axially symmetric Laplacian, indeed the same is true of the class of $AdS_5\times S^2$ solutions in IIA \cite{lm} one obtains by dimensionally reducing the M-theory class of Lin-Lunin-Maldecena \cite{llm}.  The M-theory class actually depend on a 3d Toda equation, which is equivalent to the Laplacian only when one imposes an additional $U(1)$ isometry, which one then uses to reduce to IIA. As the class of this section is in massless IIA it can be lifted to M-theory, so an obvious question poses itself: Is there a class in M-theory governed by a 3d Toda from which the backgrounds in this section descend? It would be interesting to look into this and what connection, if any, this class has to $AdS_5\times S^2$ or indeed any AdS class.

\subsubsection*{Subcase: $F_0 \neq 0$}
We expect to be able to perform a similar simplification of the sytem of PDE's for $F_0\neq 0$ case, however up to this point we have failed to do so in general. However there is a special case which is far more simple, namely $\beta=\frac{\pi}{2}$. Here \eqref{eq:lastPDE1}-\eqref{eq:lastPDE3} and \eqref{eq:lastparing} force
\beq
e^{2A}= \frac{\rho}{g(x)^{\frac{1}{4}}\sin\alpha},~~~ e^{\Phi-5A}=\frac{\cos\alpha \sin^2\alpha}{c^2\rho^2},~~~d\alpha=dc=0,
\eeq
all that remains is to ensure that $dF_0=0$ which is ensured as long as
\beq
g = \tilde c- \frac{2F_0 x\cos\alpha}{c_1^4}.
\eeq
This solution has all the generic fluxes except the internal part of $F_6$ turned on, it bears some resemblance to D8-branes on some sort of cone, but the precise picture depends on what values the free constant $\alpha$ takes.\\

We shall come back to study the solutions that follow from these massless and massive systems in \cite{DNJtocome}.

\section{The unique type II AdS$_4\times S^3$ background}\label{sec:ads4}
We have classified all Mink$_3\times S^3$ with internal Killing spinors of equal norm, up to certain PDE determining various warp factors. As equal norms is a requirement\footnote{It is established in Appendix \ref{susyconditions}  that the 7d spinors $\chi_1,~\chi_2$ must obey the relation $|\chi_1|^2\pm|\chi_2|^2=c_{\pm}e^{\pm A}$ where $c_{\pm}$ are constants. We can, without loss of generality solve these conditions in terms of unit norm spinors $\chi^0_i$ and an angle $\zeta$ as
\beq
\chi_1 = \frac{1}{\sqrt{2}} e^{\frac{A}{2}}\sqrt{1+ \sin\zeta}\chi^0_1,~~~~\chi_2 = \frac{1}{\sqrt{2}} e^{\frac{A}{2}}\sqrt{1- \sin\zeta}\chi^0_2,~~~ c_+=1,~~~c_-= e^{2A}\sin\zeta
\eeq
To make a Mink$_3$ solution AdS$_4$ requires us to fix the dependence of $e^{2A}$ on the AdS radius, but since $e^{2A}\sin\zeta$ is constant, we must either set $\zeta=c_-=0$ or fix $\zeta$ such that it also depends on the AdS radius. The latter contradicts the assumption of an $SO(2,3)$ isometry, so we conclude that $AdS_4$ requires $c_-=0$;~ ie  equal 7d spinor norms.} for the existence of AdS$_4$ it is reasonable to ask whether such solutions are contained within our classification. Any  AdS$_4$ solution can be expressed as a Mink$_3$ solution, one need only parametrise  AdS as the Poincar\'{e} patch.  This comes about quite naturally in terms of the Mink$_3\times M_7$ set up by imposing that
\beq\label{eq:ads4form}
ds^2= e^{2A}ds^2(\mathbb{R}^{1,2}) + ds^2(M_7) = e^{2\tilde{A}} \left(r^2 ds^2(\mathbb{R}^{1,2})+ \frac{dr^2}{r^2}\right)+ ds^2(M_6)~,
\eeq
with $e^{2\tilde{A}}$ and $M_6$ independent of $r$. As we have local expressions on $M_7=S^3\times M_4$, this makes our task relatively easy. A quick scan through the various cases in section \ref{IIB} and \ref{IIA} indicates that the only class that are potentially compatible with AdS$_4$ are in sections \ref{sec:IIA_a0_bpi2} and \ref{sec:novel} - the others manifestly cannot be put in the form \eqref{eq:ads4form}. These are both in IIA, but closer inspection leads one to realise that the class of section \ref{sec:novel} cannot work as \eqref{eq:BPSfinal1} would break the putative AdS isometry. This leaves only sections \ref{sec:IIA_a0_bpi2}.

We will now show that there is a unique compact\footnote{Strictly speaking section \ref{sec:first} contains AdS$_5\times S^5$ which can be expressed as a non-compact AdS$_4$ solution. We will  disregard such higher dimensional AdS solutions.} AdS$_4$ solution, at least locally, for the class of solutions of section \ref{sec:IIA_a0_bpi2}.
This background corresponds to a foliation of AdS$_4\times S^3\times S^2$ over an interval and is the near-horizon of a D2-D6 brane intersection, and can also be obtained by dimensionally reducing a certain $\mathbb{Z}_k$ orbifold of AdS$_4 \times S^7$. Starting from M-theory one first parameterises $S^7$ as a foilation of $S^3\times S^3$ over a closed interval, then performs both the orbifolding and reduction to IIA on the Hopf fibre of one of the $S^3$'s (see e.g.~\cite{Lozano:2016wrs}) - there by preserving 16 supercharges.
For this purpose, we take the metric \eqref{d2d61}, expressed in the form
\begin{equation*}
 ds^2= e^{2A} ds^2(\mathbb{R}^{1,2}) +e^{-A+\Phi}\left(d\rho^2+ \rho^2 ds^2(S^3)\right)+e^{-2A} (dx_1^2+dx_2^2+dx_3^3)\,,
\end{equation*}
and assume an AdS$_4$ factor, which requires $e^{2A}=r^2\,e^{2 \tilde{A}}$, with the rescaled warp factor $\tilde{A}$ as well as the dilaton $\Phi$ undetermined functions independent of the  AdS$_4$ radial coordinate $r$. Furthermore, the background is only $SO(2,3)$ invariant if the internal metric is independent of $r$, which fixes $\r$ and the $x_i$ to scale as $\r\sim r^{1/2}$ and $x_i\sim r$. Keeping this in mind, we parametrise
\begin{align*}
 x_1 &= r\,q(\m)\,\sin\theta\,\cos\phi \,, \nn \\
 x_2 &= r\,q(\m)\,\sin\theta\,\sin\phi \,, \nn \\
 x_3 &= r\,q(\m)\,\cos\theta \,, \nn \\
 \r  &= r^{1/2}\,h(\m) \,,
\end{align*}
where $q(\m)$ and $h(\m)$ are undetermined functions of some coordinate $\mu$, and the $(\theta,\phi)$ directions parametrise a 2-sphere such that the $\mathbb{R}^3$ spanned by $x_i$ is written in polar coordinates with radius $r\,q(\m)$.
 Now we have to ensure that the metric is diagonal with respect to the $r$-direction, i.e.~set $g_{r\,\mu}=0$, and that it shows the $1/r^2$ behaviour for $g_{rr}$, which amounts to imposing $g_{rr} = e^{2 \tilde{A}}/r^2$. These two conditions lead to the following expressions for $\tilde{A}$ and $\Phi$ in terms of $q(\m), h(\m)$ and independent of $(\theta,\phi)$:
\begin{align*}
  e^{4 \tilde{A}} &= q(\m) \left( q(\m) - h(\m)\frac{q'(\m)}{2h'(\m)} \right) \,,\\
  e^\Phi &= -2 e^{-\tilde{A}} \, \frac{q(\m)\,q'(\m)}{h(\m)\,h'(\m)} \,.
\end{align*}
These expressions imply, once inserted in the first eq.~of \eqref{eq:IIA_al0b10a11_f}, the following ODE for the $q$ and $h$ functions:
\begin{equation*}
 q'(\m) \left[ h'(\m)^2 + h(\m)\,h''(\m) \right] = h(\m)\,h'(\m)\,q''(\m) \,,
\end{equation*}
which can be solved in closed form as $h=h\big(q(\m)\big)$ and also implies the Bianchi identities of the fluxes. As $h$ is a function of $q$, rather than $\mu$, we can use diffeomophism invariance to fix $q$ such that $h$ is simple, without loss of generality we choose
\[ q(\m)= \frac{2L^3}{k} \cos^2\left(\frac{\mu}{2}\right)\,, \]
where $L$ and $k$ are constants. This leads to
\[h(\mu)=-2 L^{3/2}\sin\left(\frac{\mu}{2}\right)\,.\]
The resulting metric is of the form
\eq{
ds^2 = \frac{2 L}{k} \cos\left( \frac{\mu}{2}\right) \left[ ds^2 ( \text{AdS}_4)
+  L^2 \left( d \mu^2 + 4 \sin^2 \left( \frac{\mu}{2}\right)ds^2 (S^3) + \cos^2 \left( \frac{\mu}{2}\right) ds^2 (S^2) \right)\right]
}
with fluxes
\eq{
F_2 = -\frac{k}{2} \text{Vol}(S^2)~,~~~~ F_4 = \frac{3}{L}\text{Vol(AdS}_4)~.
}
This is the IIA reduction of AdS$_4\times S^7/\mathbb{Z}_k$ with length scale $L$ and $k$ D6-branes, as in eq.~(2.8) of \cite{Lozano:2016wrs}.

The fact that $(\theta,\phi)$ are isometry directions of this solution means that there is an additional $SU(2)_{S^2}$ symmetry due to the round $S^2$ factor in the metric and fluxes. The spinors of this solution are then charged under $SU(2)_{S^2}$ and just one of the $SU(2)$'s of $S^3$ (see the $\nu$ dependence of \eqref{d2d62}), $SU(2)_+$ say. Since $S^2$ and $S^3$ appear as a product the spinors are actually charged under  $SU(2)_+\times SU(2)_{S^2} $ which realises an enhanced $SO(4)$ R-symmetry as required by the $\mathcal{N}=4$ super-conformal algebra in 3d - $SU(2)_-$, under which the spinors are not charged, is a flavour symmetry.

\section{Type II with a single Killing spinor}\label{e20}
In the previous two sections, we have worked out the supersymmetry conditions making use of the pure spinor equations \eqref{7dsusy}, which are valid only in case $|\chi_1|^2 = |\chi_2|^2$. Note that this is a necessary condition for the existence of D-branes which do not break background supersymmetry.The supersymmetry condition for a $\text{D}_p$-brane is given by $\G^{(p)} \e_1 = \e_2$. Since $\G^{(p)}$ is unitary, squaring this equation leads to the conclusion that left- and right-handside must have equal norm.\footnote{The argument is slightly more complicated due to the fact that $Spin(1,9)$ spinors do not admit a non-trivial norm, and hence one should decompose to $Spin(9)$ first. See \cite{kt} for details. Also note that strictly speaking, the norms need only be equivalent on the brane.} We will examine the simplest non-equal norm case, namely the one where
\eq{
\e_2 = 0~.
}
We could either make use of the generalised geometrical reformulation of supersymmetry which incorporates $|\chi_1|^2 - |\chi_2|^2 \neq 0$ as deduced in appendix \ref{susyconditions}, or use the actual Killing spinor equations. Considering the simplicity of this case, we will use the latter.

Much of the work has however already been done: the conditions for seven-dimensional pure NSNS solutions have been deduced in \cite{gkmw, friedrich1, friedrich2} up to some ans\"{a}tze. We will merely show that the ans\"{a}tze made in \cite{gkmw, friedrich1, friedrich2} (no warp factor, no external NSNS flux) are in fact enforced by supersymetry, and then proceed to plug in the decomposition resulting from $M_7 = S^3 \times M_4$. This leads to a pair of explicit pure NS backgrounds: the NS5-brane and the U-dual to the IIB conical backgrounds of section \ref{sec: additional S3}. We will also analyse the seven-dimensional RR-sector, which is new, but the conclusion is that all RR-fluxes vanish.

\subsection{Seven-dimensional decomposition}
Our starting point are the democratic supersymmetry equations, which read as follows for $\e_2 = 0$:
\eq{\label{10dsusy}
\left( \slashed{\p} \phi - \frac12 \slashed{H} \right) \e_1 = \left(\nabla_M - \frac14 \slashed{H}_M \right) \e_1 &= 0~,~~~~
\l \slashed{F} \G_M \e_1 = 0 ~.
}
As can be seen, the NSNS and RR sectors decouple. We impose a similar $3+7$ decomposition as before: the metric and RR flux is given by \eqref{eq:main_Ansatz}, while the Killing spinor $\e_1$ is given by \eqref{10dks} and $\e_2 = 0$. We generalize the NSNS 3-form flux by allowing a term $h~ e^{3A} \text{Vol}_3$.

Using the convention $\g_{\m\n\r} =  \e_{\m\n\r}$, plugging the above decompositions into \eqref{10dsusy} leads to the following 7d equations:
\begin{subequations}\label{7dsusy2}
\begin{align}
\left( \p_m \phi \g^m -\frac{1}{12} H_{mnp} \g^{mnp} + \frac12 i h \right) \chi =
\left( \nabla_m^{(7)} - \frac18 H_{mnp} \g^{np} \right) \chi &= 0 \label{7dsusya}\\
\left( \frac14 e^A h - i e^A \p_m A \g^m \right) \chi &= 0 \label{7dsusyb}\\
\l f \chi = \l f \g_m \chi &= 0 ~.\label{7dsusyc}
\end{align}
\end{subequations}
From \eqref{7dsusyb} it follows that
\beq
\partial_m A= h =0 \;.
\eeq
As can be seen, the NSNS and RR sector split and can thus be analysed independently.

The existence  of a globally defined nowhere-vanishing $Spin(7)$ Majorana spinor $\chi$ reduces the structure group of $M_7$ to $G_2$. More concretely, the following bilinears can be defined:
\beq
\varphi_{mnp}=- i\chi^{\dag}\gamma_{mnp}\chi~,~~~~ (\star_7 \varphi)_{mnpq}=   \chi^{\dag}\gamma_{mnpq}\chi ~,
\eeq
where we have normalized $\chi$. The other bilinears, i.e., the 1-, 2-, 5- and 6-form vanish.  As has been deduced in \cite{gkmw, friedrich1, friedrich2}, \eqref{7dsusya} can be rewritten in terms of the $G_2$-structure as
\beq\label{eq: G2susy}
d \varphi \wedge \varphi = d(e^{-2\Phi}\star_7\varphi )=d(e^{-2\Phi}\varphi) - e^{-2\Phi}\star_7 H=0~.
\eeq
We will analyse the solution to the NSNS sector by requiring a further splitting of $M_7 = S^3 \times M_4$. On the other hand, we will show that the RR-fluxes vanish for any $M_7$.

\subsection{NSNS sector}
Considering the case $M_7 = S^3 \times M_4$, we further decompose the spinor as
\beq
\chi= \xi\otimes\left(\sin(\alpha/2) \eta+ \cos(\alpha/2) \hat\gamma\eta\right) + \text{m.c.}, ~~~~\eta^{\dag}\eta=1,~~~~\eta^{\dag}\hat\gamma\eta=0 ~,
\eeq
with m.c. the Majorana conjugate.
This leads to a further reduction of the structure group. Since $S^3$ is parallelisable, it has trivial structure group, leading to a $Spin(4)$ structure group on $M_4$. Generically, the structure group need not reduce on $M_4$.\footnote{Note that on any $M_7$ with a spin structure, an $SU(2)$-structure can be found \cite{fkms} \cite{kmt}. However, this is not necessarily the structure group defined by the spinors we are making use of, and so even when splitting $M_7 = M_3 \times M_4$ with $M_3$ parallelisable, $M_4$ need not admit a globally well-defined $SU(2)$-structure; consider for example $M_7 = S^3 \times S^4$.}
In the case where either $\eta_+$ or $\eta_-$ is nowhere vanishing, the structure group reduces to $SU(2)$, in case both are nowhere-vanishing, the structure group is trivial. As everywhere else, our analysis is purely local and we will work with a local trivial structure, parametrising possible vanishing of either chiral spinor by the angle $\a$.\\

First, as in  \cite{Gaillard:2010gy}, we make use of an auxiliary $SU(3)$-structure $(J, \O)$ to express the $G_2$-structure as
\beq
\varphi = - v_2\wedge J - \text{Im}\Omega,~~~~\star_7 \varphi= \frac{1}{2}J\wedge J+ \text{Re}\Omega\wedge v_2 ;.
\eeq
Next, we decompose the $SU(3)$-structure in terms of the vielbeine as
\beq
J=-\frac{1}{2}(K_1\wedge w_1+ K_2\wedge w_2+ K_3\wedge v_1),~~~\Omega =e^{i \alpha}(\frac{K_1}{2}+ i w_1)\wedge(\frac{K_2}{2}+ i w_2)\wedge(\frac{K_3}{2}+ i v_1)
\eeq
Inserting this into \eqref {eq: G2susy}, one finds
\begin{subequations}
\begin{align}\label{eq:G2S3eqs}
&d(e^{3C-2\Phi}\cos\alpha v_2)= d(e^{2C-2\Phi} u_i)+\nu e^{C-2\Phi}(2u_i\wedge v_2+ \sin\alpha\epsilon_{ijk}u_j\wedge u_k)=0~,\\[2mm]
&d(e^{2C-2\Phi}(\sin\alpha u_i\wedge v_2+ \frac{1}{2}\epsilon_{ijk}u_j\wedge u_k))-\nu\epsilon_{ijk} e^{C-2\Phi}\cos\alpha u_j\wedge u_k\wedge v_2=0~,\\[2mm]
&d\alpha\wedge v_1\wedge w_1\wedge w_2=0~,~~~~ u=(v_1,w_1,w_2),\\[2mm]
&e^{-2\Phi}\star_7H= -d(e^{-2\Phi} \cos\alpha v_1\wedge w_1\wedge w_2)-\nu \text{Vol}(S^3)\wedge d(e^{3C-2\Phi}\sin\alpha).
\end{align}
\end{subequations}
When $\alpha\neq\frac{\pi}{2}$,  by taking linear combinations, exterior derivatives and wedge products with the vielbein of the equations in \eqref{eq: G2susy}, one can derive
\beq
\epsilon_{ijk}\big[d(e^{C-\Phi}\cos\alpha)-\nu e^{-\Phi}v_2\big]\wedge u_j\wedge u_k+ e^{5C+12\Delta}\cos^4\alpha\big[d(e^{-3C+2\Phi}\sec\alpha\tan\alpha)\wedge v_2]\wedge u_i=0\nn,
\eeq
where, since $u_i$ form a basis of independent 1-forms, the terms in square parentheses must vanish.  This is sufficient to conclude that
\beq
d\alpha=0~,~~~ C= C(\rho)~,~~~\Phi= \Phi(\rho)~,~~~~\rho= e^{C-\Phi}~.
\eeq
It is then not hard to establish that
\beq
du_i\wedge u_j=0~,~~~i\neq j,
\eeq
in a similar fashion. This means we can locally parametrise
\beq
v_2= \nu\sec\alpha e^{\Phi} d\rho~,~~~  du_i=c_i \epsilon_{ijk}u_j\wedge u_k~,~~~dc_i=0~.
\eeq
Plugging this back into \eqref{eq: G2susy} we find that
\beq
c_i= e^{-C}\nu\tan\alpha,
\eeq
so either $dC=0$ or $c=\alpha=0$.\\

\textbf{Case 1}:
When $\alpha=0$, $u_i$ are the vielbeine of $T^3$ so we can simply take
\beq
u_i= dx_i.
\eeq
All that is left to do is calculate $H$ and impose its Bianchi identity. We find
\beq\label{ns52}
H= \nu \partial_{\rho}(e^{2\Phi})\rho^3\text{Vol}(S^3) ~.
\eeq
Closure of the flux then implies that $e^{2 \Phi}$ is harmonic, leading to
\beq\label{ns53}
e^{2\Phi}=g_s^2 \left(1 + \frac{c}{\rho^2}\right)~,~~~~~~~ g_s,c \in \rbb~,
\eeq
which is consistent with the definition of $\rho$.
Finally, we note that the metric is given by
\beq\label{ns51}
ds^2= ds^2(\rbb^{1,5})+ e^{2\Phi}\left(d\rho^2+ \rho^2 ds^2(S^3)\right)~.
\eeq
This is an NS5-brane, dual to the D5-brane solution in section \ref{eq:D5case} \cite{johnson}.\\

\textbf{Case 2}:
When $0<\alpha<\frac{\pi}{2}$, $u_i$ span the vielbeine of another $S^3$ so we take
\beq\label{consistency1}
u_i = \frac{e^{\tilde{C}}}{2}\tilde{K}_i,~~~d\tilde{K}_i+\frac{\tilde{\nu}}{2}\epsilon_{ijk}d\tilde{K}_j\wedge d\tilde{K}_k,~~~d\tilde{C}=0~,
\eeq
where consistency requires that
\beq\label{consistency2}
\tilde{\nu} e^{C}\cos\alpha+ e^{\tilde{C}}\nu\sin\alpha=0~,
\eeq
and find
\eq{\label{e20s3s32}
H=  - 2 e^{3C}\partial_{\rho}(e^{-\Phi})\cos^2\alpha  \text{Vol}(S^3)
 - 2 e^{3\tilde{C}}\partial_{\rho}(e^{-\Phi})\cos\a\sin\alpha \text{Vol}(\tilde{S}^3) ~.
}
By definition of $\rho$, the Bianchi identity is satisfied. After redefining $r = \exp\left( e^{-C} \cos \a \rho\right)$, it follows that the metric is given by
\eq{\label{e20s3s310}
ds^2= ds^2(\rbb^{1,2})+ d r^2 + e^{2C}ds^2(S^3)+  \cot^2 \a e^{2C}  ds^2(\tilde{S}^3)~.
}
Note that in IIB, this solution can be obtained from the solution of section \ref{sec: additional S3} by means of the following S-duality transformation (up to redefining some constants):
\beq\label{e20s3s31}
\Phi \rightarrow - \Phi~, ~~~~ ds^2 \rightarrow e^{-\Phi} ds^2~,~~~~F_3 \rightarrow - H~.
\eeq

\subsection{RR-sector}
The NSNS sector has been analysed by imposing a further decomposition $Spin(7) \rightarrow Spin(3) \times Spin(4)$ on the spinor. On the other hand, we will analyse the RR-sector in full generality.

Let us consider the RR-flux constraints equations \eqref{7dsusyc}, repeated here for convenience:
\eq{\label{7dsusyrr}
\l \slashed{f} \chi = \l \slashed{f} \g_m \chi = 0 ~.
}
For type IIA, we have that
\eq{
\l \slashed{f} = f_0 + - \slashed{f_2} - i \slashed{f_3} - i \slashed{f_1} ~,
}
where we have defined $f_3 = \star_7 f_4$, $f_1 = \star_7 f_6$.
For type IIB, one finds
\eq{
\l \slashed{f} = \slashed{f_1} - \slashed{f_3} - i \slashed{f_2} - i f_0 ~,
}
with $f_2 = \star_7 f_5$, $f_0 = \star_7 f_7$, hence up to some field redefinitions, the supersymmetry constraints are identical.
The fluxes, a priori irreducible representations of $SO(7)$, decompose into representations of $G_2$ as follows:
${\bf 7 } \rightarrow {\bf 7}$, ${\bf 21} \rightarrow {\bf 7} + {\bf 14}$, ${\bf 35}
\rightarrow {\bf 1} + {\bf 7} + {\bf 27}$. Concretely, we parametrise
\eq{
f_{2|mn} &= \varphi_{mnp} f_2^p + f_{2|mn} \\
f_{3|mnp} &= f_3 \varphi_{mnp} + \psi_{mnpq} f_3^q + f_{3|q[m}
\varphi_{np]}^{\phantom{np]}q} ~,
}
where the ${\bf 14}$ satisfies $f_{2|mn} = \frac12 \psi_{mnpq} f_2^{pq}$ and the
{\bf 27} is corresponds to a symmetric traceless 2-tensor. Furthermore, we have introduced the notation $\psi = \star_7 \varphi$ for convenience.
Making use of the $G_2$-structure identities \eqref{g21}, \eqref{g22}, we find
\eq{
\begin{alignedat}{6}
\slashed{f_1} \chi &= f_{1|m} \g^m \chi                  & \qquad & \slashed{f_1} \g_m \chi &=&     f_{1|m}\chi - i \varphi_{mnp} f_1^n \g^p \chi \\
\slashed{f_2} \chi &= 3 i f_{2|m} \g^m \chi              & \qquad & \slashed{f_2} \g_m \chi &=& 3 i f_{2|m}\chi -   \varphi_{mnp} f_2^n \g^p \chi  - 2 f_{2|mn}\g^n \chi \\
\slashed{f_3} \chi &= 7 i f_3 \chi - 4 f_{3|m} \g^m \chi & \qquad & \slashed{f_3} \g_m \chi &=& - i f_3\g_m  \chi + 4 f_{3|m} \chi + 6 i f_{3|mn} \g^n \chi ~.
\end{alignedat}
}
Inserting the above into \eqref{7dsusyrr} and comparing representation by representation, it follows that all RR fluxes vanish.

\section*{Acknowledgments}
We would like to thank Alessandro Tomasiello for collaboration in the early stages of this project. D.P. would like to thank Noppadol Mekareeya for discussion. J.M. is grateful to the Physics Department of Milano-Bicocca U., the ITP at KU Leuven and SISSA for the warm hospitality. Throughout this project N.M. has been variously funded by INFN; the European Research Council under the European Union's Seventh Framework Program (FP/2007-2013); ERC Grant Agreement n. 307286 (XD-STRING); and the Italian Ministry of Education , Universities and Research under the  Prin project "Non-Perturbative Aspects of Gauge Theories and Strings" (2015MP2CX4). J.M. is supported by the FPI grant BES-2013-064815 of the Spanish MINECO, and the travel grant EEBB-I-17-12390 of the same institution. He also acknowledges partial support through the Spanish and Regional Government Research Grants FPA2015-63667-P and FC-15-GRUPIN-14-108. D.P. is funded by  ERC Grant Agreement n. 307286 (XD-STRING).

\appendix
\section{Conventions \& identities}\label{conv}

We decompose ten-dimensional gamma matrices as
\eq{
\G_\m = \s_3 \otimes e^A \g_\m \otimes \obb ~,~~~~
\G_m = \s_1 \otimes \obb \otimes \g_m ~,
}
and seven-dimensional gamma matrices as
\beq\label{gammadecomp}
\gamma^{(7)}_\a = e^{C}\sigma_\a\otimes \hat\gamma~,~~~
\gamma^{(7)}_a = \mathbb{I}\otimes \gamma_a,~~~B_7= \sigma_2\otimes B_4,~~~ B_4B_4^*=-\mathbb{I} ~.
\eeq

\subsection{$M_7$}
We consider gamma matrices satisfying
\eq{
\g_{mnpqrst} &=  i \e_{mnpqrst}~,~~~~
\g_{(7-n)} = (-1)^{\frac12 n (n-1)} i \star_7 \g_{(n)} ~.
}
A nowhere-vanishing $Spin(7)$ spinor defines a $G_2$-structure on $M_7$ by means of the bilinear
\eq{
\varphi_{mnp} = -i \chi^\dagger \g_{mnp} \chi ~.
}
Defining $\psi = \star_7 \phi = \chi^\dagger \g_{mnpq} \chi$, the $G_2$-structure satisfies the following identities \cite{bds}:
\eq{\label{g21}
\psi_{mnrs} \psi_{pq}^{\phantom{pq}rs} &= - 2 \psi_{mnpq} + 4 \delta_{mp} \delta_{nq} - 4 \delta_{mq} \delta_{np}~,  \\
\psi_{mnrs} \varphi_{p}^{\phantom{p}rs} &= - 4 \varphi_{mnp}~, \\
\varphi_{mrs} \varphi^{nrs} &= 6 \delta_m^n ~,
}
as well as
\eq{\label{g22}
\g_{mn} \chi  &= i \varphi_{mnp} \g^p \chi~, \\
\g_{mnp} \chi &= i \varphi_{mnp} \chi - \psi_{mnpq} \g^q \chi~, \\
\g_{mnpq} \chi &= - 4 i \varphi_{[mnp} \g_{q]} \chi + \psi_{mnpq} \chi ~.
}

\subsection{$S^3$}
We consider Pauli-matrices playing the role of gamma-matrices. They satisfy
\eq{
\s_{\a\b\g} &=  i \e_{\a\b\g}~,~~~~
\s_{(3-n)} = (-1)^{\frac12 n (n-1)} \star_3 \s_{(n)} ~,
}
with $\a,\b,\g = 1,2,3$ indices on $S^3$. The non-vanishing spinor bilinears of $S^3$ are given by
\eq{\label{s3bilin}
\xi^{c\dagger} \s^\a \xi  &= \frac12 \left(K_1^\a + i K_2^\a \right) ~,~~~
\xi^\dagger \s^\a \xi = \frac12 K_3^\a ~.
}
The real 1-forms $K_i$, $i = 1,2,3$, define a trivial structure on $S^3$ (i.e., a vielbein, up to normalisation). Note that $S^3$ is parallelisable and hence the trivial structure is globally well-defined. We will always normalise the volume form as
\beq
K_1\wedge K_2\wedge K_3=-8\text{Vol}(S^3)~,
\eeq
regardless of which specific vielbein is used. \\

\section{The bispinors of $M_4$}\label{4dbispinors}
We consider gamma matrices satisfying
\eq{
\g_{abcd} &= \e_{abcd} ~,~~~ \g_{(4-n)} = (-1)^{\frac12 n (n+1)} \star_4 \g_{(n)} \hat{\g} ~,
}
with $\hat{\g} = \g_{1234}$ the chirality matrix. Given a globally well-defined nowhere vanishing chiral spinor $\eta_+$, one can construct the bilinears
\eq{
J_{ab} = i \eta^\dagger_+ \g_{ab} \eta_+ ~, ~~~~ \o_{ab} = i \eta^{c\dagger}_+ \g_{ab} \eta_+
}
which furnish an $SU(2)$-structure. Given two globally well-defined nowhere vanishing chiral spinors of opposite chirality, the structure group reduces to a trivial structure \cite{lpt}. Generically, supersymmetry requires a nowhere vanishing spinor $\eta$, which can admit a chiral locus. This ensures that, although the structure group of $M_4$ cannot be globally reduced, it is possible to reduce the structure group of the generalised cotangent bundle $T M_4 \oplus T^* M_4$ to $SU(2) \times SU(2)$, completely analogously to the well-known situation of $SU(3)$-structures \cite{gmpt}. Since the supersymmetry constraints are local, we will always work with the vielbeine determining the local trivial structure. Using the conventions of \cite{Apruzzi:2014qva} with $\eta = (\eta_+, \eta_-)$, we set
\beq\label{eq: canonicalframe}
v=v_1+ i v_2 = \eta^{\dag}_- \gamma_a \eta_+ dx^a, \quad w= w_1+ i w_2 =\eta^{c\dag}_- \gamma_a \eta_+ dx^a~.
\eeq
Although some care must be taken on the chiral locus, where the above 1-forms all vanish, it turns out that no solutions exist on the chiral locus, as discussed in sections \ref{IIB} and \ref{IIA}.
We can expand the locally defined 4d components of the Killing spinors $\eta_{1,2}$ in terms of $\eta$ as
\beq
\eta^{\dag}\eta=1~,~~~\eta^{\dag}\hat\gamma\eta=0
\eeq
as
\beq\label{eq: general4dspinors}
\eta_1= \cos\left(\frac{\alpha}{2}\right) \eta+ \sin\left(\frac{\alpha}{2}\right)\hat\gamma \eta~,~~~\eta_2= a\eta+ b \hat\gamma\eta+ c \eta^c+ d \hat\gamma \eta^c
\eeq
where $a,b,c$ and $d$ are subject to
\beq\label{eq:sum of squares}
|a|^2+ |b|^2+ |c|^2+ |d|^2=1~.
\eeq
We can then calculate the 4d bispinors appearing in \eqref{eq: IIBSUSYa}-\eqref{eq: IIBSUSYf} and \eqref{eq: IIaSUSYa}-\eqref{eq: IIASUSYf}, where to do so we find it useful to parametrise
\beq
a=a_1+i a_2~,~~~ b=b_1+i b_2~,~~~c=c_1+i c_2~,~~~ d=d_1+i d_2~,
\eeq
for $a_i,b_i,c_i$ and $d_i$ real. However we first note that in both IIA and IIB we must solve the 0-form constraints
\beq
(\psi^2_{\hat\gamma})_0 = (\text{Im}\psi^1_{\hat\gamma})_0=0~,
\eeq
which reduce to
\beq
b_2 \cos\frac{\alpha}{2}+ a_2 \sin\frac{\alpha}{2}=d_2 \cos\frac{\alpha}{2}+ c_2 \sin\frac{\alpha}{2}=d_1 \cos\frac{\alpha}{2}+ c_1 \sin\frac{\alpha}{2}=0~.
\eeq
We can solve these in general by fixing
\eq{
\begin{alignedat}{8}
a_2&= \l_1\cos(\frac{\alpha}{2}) &\qquad&  b_2&{}={}& -\l_1\sin(\frac{\alpha}{2}) &\qquad&   c_1&{}={}&  \l_2\cos(\frac{\alpha}{2})~,\\[2mm]
c_2&=-\l_3\cos(\frac{\alpha}{2}) &\qquad&  d_1&{}={}& -\l_3\sin(\frac{\alpha}{2}) &\qquad&   d_2&{}={}& -\l_1\sin(\frac{\alpha}{2})~,
\end{alignedat}
}
which turns \eqref{eq:sum of squares} into
\beq
a_1^2+ b_1^2+ \l_1^2+ \l_2^2+\l_3^2=1.
\eeq
In terms of this parametrisation the 4d bispinors are given by
\begin{align}\label{eq:4dbispinors}
 \psi^1_{+}
 &=  a_1 - i\,\l_1 -i\,b_1v_1\wedge v_2  -  (\l_2-i\l_3)v_1\wedge (w_1-i\,w_2)  +  (i\,a_1+\l_1) w_1\wedge w_2 \nn \\
 & ~~~~~~   +  b_1v_1\wedge v_2\wedge w_1\wedge w_2  \,, \nn \\[2mm]
 \psi^1_{-}
 &=   (a_1-i\,\l_1)v_1  -  i\,b_1 v_2  -  (\l_2-i\,\l_3)(w_1-i\,w_2)  + (i\,a_1+\l_1) v_1\wedge w_1\wedge w_2  \nn \\
 & ~~~~~~   +  b_1 v_2\wedge w_1\wedge w_2 \,, \nn \\[2mm]
 \psi^2_{+}
  &=  -(\l_2+i\l_3) - (a_1+i\,\l_1)v_1\wedge (w_1-i\,w_2)  -  i\,b_1 v_2\wedge (w_1-i\,w_2) \nn \\
  & ~~~~~~  -  i(\l_2+i\,\l_3)w_1\wedge w_2  \,, \nn \\[2mm]
 \psi^2_{-}
 &= -  (\l_2+i\,\l_3)v_1  -  (a_1+i\,\l_1)(w_1-i\,w_2)  -  i\,b_1 v_1\wedge v_2\wedge (w_1-i\,w_2)   \nn \\
 &~~~~~~  +  (\l_3-i\,\l_2) v_1\wedge w_1\wedge w_2   \,, \\[2mm]
 \psi^1_{\hat\gamma+}
 &= b_1  -  (i\,a_1+\l_1)v_1\wedge v_2  -  i\,(\l_2-i\,\l_3)v_2\wedge (w_1-i\,w_2)  +  i\,b_1w_1\wedge w_2 \nn\\
 &~~~~~~   +  (a_1-i\,\l_1)v_1\wedge v_2\wedge w_1\wedge w_2  \,, \nn \\[2mm]
 \psi^1_{\hat\gamma-}
 &= -  b_1 v_1  +  (i\,a_1+\l_1)v_2  +  (\l_3+i\,\l_2) v_1\wedge v_2\wedge (w_1-i\,w_2)  -  i\,b_1 v_1\wedge w_1\wedge w_2 \nn \\
 &~~~~~~   -  (a_1-i\,\l_1) v_2\wedge w_1\wedge w_2 \,,\nn \\[2mm]
 \psi^2_{\hat\gamma+}
 &= (i\,\l_2-\l_3) v_1\wedge v_2  -  b_1 v_1\wedge (w_1-i\,w_2)  -  i(a_1+i\,\l_1)v_2\wedge (w_1-i\,w_2)    \nn\\
 &~~~~~~  -  (\l_2+i\,\l_3) v_1\wedge v_2\wedge w_1\wedge w_2  \,, \nn \\[2mm]
 \psi^2_{\hat\gamma-}
  &= (\l_3-i\,\l_2)v_2  +  b_1 (w_1 - i\,w_2)  +  i(a_1+i\l_1) v_1\wedge v_2\wedge (w_1-i\,w_2)  \nn\\
  & ~~~~~~   +  (\l_2+i\,\l_3) v_2\wedge w_1\wedge w_2 \,. \nn
\end{align}

\section{The $SU(2)$ doublets of $S^3$}\label{sec: su2doublet}
There exist two independent spinors on $S^3$ that obey the Killing spinor relations
\beq\label{eq:S3KSE}
\nabla_a \xi_{\pm}=\pm\frac{i}{2} \gamma_a\xi_{\pm},
\eeq
each of which preserves two supercharges. Additionally the global isometry group of $S^3$ can be decomposed as $SO(4)=SU(2)_+\times SU(2)_-$, so $S^3$ supports two sets of $SU(2)$ Killing vectors  $K^i_{\pm}$ , $i=1,2,3$, that are dual to one forms that obey
\beq
dK_i^{\pm} \pm \frac{1}{2}\epsilon_{ijk} K_j^{\pm}\wedge K_k^{\pm},
\eeq
i.e. the  right-/left-invariant forms of $SU(2)$. It is possible to use the spinors on $S^3$ to construct $SU(2)_{\pm}$ doublets. Consider the following vector with spinor entries
\beq
\xi^{\a}_{\pm}= \left(\begin{array}{c}\xi_{\pm}\\\xi^c_{\pm}\end{array}\right)^{\a}.
\eeq
These transform under the action of the spinoral Lie derivative as\footnote{The spinoral lie derivative along a Killing vector $K$ is defined as
\beq
\mathcal{L}_{K}\epsilon= K^{\mu}\nabla_{\mu}\epsilon + \frac{1}{8}(dK)_{\mu\nu}\gamma^{\mu\nu}\epsilon.
\eeq
The easiest way to see that this leads to the claimed transformation property, is to parametrise the vielbein on $S^3$ as $e^1=\frac{1}{2}d\theta,~~e^2=\frac{1}{2}\sin\theta d\phi,~~e^3=\frac{1}{2}(d\psi+ \cos\theta d\phi)$ and take the flat space gamma-matrices to be the Pauli matrices $\sigma_i$. Then \eqref{eq:S3KSE} is solved by $\xi_{+}= e^{\frac{i}{2}\theta\s_1}e^{\frac{i}{2}\phi\s_3}\xi^0_+$, $\xi_{-}= e^{-\frac{i}{2}\psi\s_3}\xi^0_-$ for $\xi^0_{\pm}$ constant 2d spinors. The $SU(2)_{\pm}$ forms are then precisely $K^+_i=-i\text{Tr}(\sigma_i dg g^{-1})$, $K^-_i=-i\text{Tr}(\sigma_i g^{-1}dg )$  for $g= e^{\frac{i}{2}\phi\sigma_3}e^{\frac{i}{2}\theta\sigma_2}e^{\frac{i}{2}\psi\sigma_3}$. The result is then not hard to show.}
\beq
\mathcal{L}_{K_i^{\pm}} \xi^{a}_{\pm} =\pm\frac{i}{2}(\sigma_i)^{a}_{\phantom{a}b}\xi^{\b}_{\pm},~~~\mathcal{L}_{K_i^{\pm}} \xi^{a}_{\mp} =0,
\eeq
for $\sigma_i$ the Pauli matrices, which means that $\xi^{a}_{\pm}$ transforms as a doublet under local $SU(2)_{\pm}$ transformations  and a singlet under $SU(2)_{\mp}$.

\section{Supersymmetry conditions for three-dimensional external spacetimes}\label{susyconditions}
In \cite{hlmt}, supersymmetry conditions for 3+7 dimensional compactifications are given in terms of bispinors. The repackaging of the supersymmetry conditions was done under the following conditions:
\begin{itemize}
\item The external space is Minkowski.
\item The spinors have equivalent length.
\item The NSNS flux $H$ does not have an external component.
\end{itemize}
In this section, we will look at relaxing the latter two conditions to obtain more general solutions. Our starting point will be the ten-dimensional bispinor description of the supersymmetry constraints, as described in \cite{tom}:
\begin{subequations} \begin{align}\label{10d1}
\d_H \left( e^{-\Phi} \Psi \right)  + \tilde{K} \wedge F + \iota_K F &= 0 \\ \label{10d2}
\d \tilde{K} &= \iota_K H \\\label{10d3}
\mathcal{L}_K g_{10} &= 0 \\
\left( e_{+1} \cdot \Psi \cdot e_{+2}, \G^{MN} \left(\pm d_H \left( e^{-\Phi} \Psi \cdot e_{+2} \right) + \frac12 e^{\Phi} d^\dagger \left( e^{- 2 \Phi} e_{+2} \right)\Psi - F \right)\right) &= 0~, \\
\left( e_{+1} \cdot \Psi \cdot e_{+2}, \left(  d_H \left( e^{-\Phi} \Psi \cdot e_{+1} \right) - \frac12 e^{\Phi} d^\dagger \left( e^{- 2 \Phi} e_{+1} \right)\Psi - F \right)\G^{MN}\right) &= 0~,
\end{align}
\end{subequations}
with
\eq{
\Psi        &= \e_1 \otimes \bar{\e_2} ~,~~~
K_M         = \frac{1}{32} \left(\bar{\e_1} \G_M \e_1 +  \bar{\e_2} \G_M \e_2\right) ~,~~~~
\tilde{K}_M = \frac{1}{32} \left(\bar{\e_1} \G_M \e_1 - \bar{\e_2} \G_M \e_2\right) ~.
}
The final two equations are known as the pairing equation; we refer to \cite{tom} for more details, and will follow along the lines of section (4.2) in the following.\\

We consider the case where the Killing spinors are given by \eqref{10dks}. Due to the properties of $Spin(1,2)$, we can define
\eq{
\frac{1}{2}\bar{\zeta} \g_\m \zeta     &= v_\m ~,~~~
\frac{1}{2}\bar{\zeta} \g_{\m\n} \zeta = (\star_3 v)_{\m\n} ~,
}
with the other bilinears vanishing. Since we are considering flat space, $\z$ are covariantly constant, hence $d v = 0$.
Making use of the spinor decomposition, it follows that
\begin{align}
8 e^{-A} \Psi     &= v \wedge \Phi_\mp - \star_3 v\wedge \Phi_\pm ~,~~~~
K         = \frac18 e^{A}(|\chi_1|^2+|\chi_2|^2)v ~,~~~
\tilde{K} = \frac18 e^{A}(|\chi_1|^2-|\chi_2|^2)v,
\end{align}
where we have defined $\Phi_+ + i \Phi_- = 8 e^{-A} \chi_1 \otimes \chi_2^\dagger$ and $K,\tilde K$ should be read as 1-forms in ten dimensions. Using the flux decomposition
\eq{
F &= f + e^{3A} \text{Vol}_3 \wedge \star_7 \lambda (f) ~,~~~~
H  = H_3 + e^{3A} h \text{Vol}_3 ~.
}
We will first solve \eqref{10d3}: since by construction, $v, K$ are Killing vectors, we  must have
\beq
|\chi_1|^2+|\chi_2|^2=c_+e^{A}.
\eeq
Next lets consider \eqref{10d2}, which leads to
\eq{
c_+ e^{3A} h= 0~,~~~ |\chi_1|^2-|\chi_2|^2=c_-e^{-A} ~.
}
Next, let us consider \eqref{10d1}. We find that
\eq{
\d_{H_3} \left( e^{3A - \Phi} \Phi_\pm \right) &= c_+ e^{3A} \star_7 \lambda (f) \\
\d_{H_3} \left( e^{2A - \Phi} \Phi_\mp \right) &= c_-  f ~.
}
In addition, the fact that $c_- \neq 0$ does not change the argument of \cite{tom}, so the pairing equations remain unchanged, leading to
\eq{
(f,\Phi_\mp) = 0 ~.
}

\section{M-theory}
The focus of this paper are backgrounds in type IIA and type IIB. In this appendix, we will discuss M-theory backgrounds on $\rbb^{1,2} \times S^3 \times M_5$. Given equivalent internal spinor norms, our (massive) IIA classification is complete (up to finding solutions to PDE). Therefore, a significant number of backgrounds one would obtain from a similar analysis of M-theory are those which one can obtain from uplifting our massless IIA backgrounds. Novel solutions from a complete M-theory analysis would be backgrounds satisfying one of the two conditions: either $M_5$ does not admit an $S^1$ factor to be integrated out to perform the dimensional reduction to IIA, or the internal component of the Killing spinor on $M_8 = S^3 \times M_5$ is such that after the reduction, the resulting seven-dimensional internal components of the IIA spinors are not of equal norm. \\

Such a full M-theory classification is beyond the scope of this paper. Instead, we aim to make contact with the literature of M-theory on $\rbb^{1,2} \times M_8$, which is much studied (see for example \cite{bb, becker, sethi, Martelli:2003ki, tsimpis, pt}).
We will derive the decomposed supersymmetry conditions, and give several simple classes of solutions. \\

For $\mathcal{N} = 1$ solutions to the supersymmetry constraints on $\rbb^{1,2} \times M_8$, the Killing spinor $\e$ decomposes as
\eq{
\e = \xi \otimes \left(\chi_+ + \chi_- \right) \;,
}
where $\xi$ is a Majorana spinor of $Spin(1,2)$  and $\chi_\pm$ are chiral Majorana spinors of $Spin(8)$.
Generically, $\chi_\pm$ can have zeroes, and the structure group of $M_8$ is $SO(8)$, although a $Spin(7)$-structure can be defined on the auxiliary space $M_8 \times S^1$ \cite{tsimpis}. In the case where one of the two does not vanish, the structure group reduces to $Spin(7)$ \cite{Martelli:2003ki}. If both chiral spinors have no zeroes, both of them define a $Spin(7)$-structure: the intersection of the two leads to a reduction of the structure group to $G_2$.\footnote{Note that this is only the case for two $Spin(7)$-structures defined in terms of opposite chirality spinors. Given two same chirality globally well-defined nowhere vanishing spinors, the structure group instead reduces to $Spin(6) \simeq SU(4)$.} The reduction of the structure group leads to the existence of globally defined invariant tensors. Instead, we will work locally, and consider patches where either one or both are non-zero.

\subsection{$Spin(7)$ holonomy}\label{chiralm}
Let us first examine the case with
\eq{
\e = \zeta \otimes \chi_+ ~.
}
Following the conventions of \cite{Martelli:2003ki}, the general solution to the M-theory supersymmetry constraints with these ans\"{a}tze is that
\begin{align}
ds^2= e^{2\Delta}ds^2(\mathbb{R}^{1,2})+ e^{-\Delta}ds^2(M_8)~,~~~ G= \text{Vol}_3\wedge d(e^{3\Delta})+ F~,
\end{align}
where $ds^2(M_8)$ a metric of $Spin(7)$ holonomy. The four-form $F$ lies in the {\bf 27} of $Spin(7)$, i.e., it satisfies
\eq{\label{27}
F^{pqr}_{\phantom{pqr}m} \Psi_{npqr}= 0~,
}
with $\Psi_{mnpq} = \bar{\chi} \g_{mnpq} \chi $ the invariant four-form defining the $Spin(7)$-structure. In addition, the Bianchi identity and equation of motion for $F$ require that $F$ is harmonic and satisfies
\beq\label{tadpole}
d(\star_8 d(e^{-3\Delta}))+ \frac{1}{2} F\wedge F =0
\eeq
away from M2-brane sources.\\

Let us now impose $M_8 = S^3\times M_5$. The internal metric and Killing spinor decompose as
\beq
ds^2(M_8)= e^{2C}ds^2(S^3)+ ds^2(M_5)~,~~~
\chi_+ = \left(\begin{array}{c}1\\0\end{array}\right)\otimes(\xi \otimes \eta + \xi^c \otimes \eta^c)~,
\eeq
and the gamma matrices decompose as
\beq\label{eq:gammasforMtheory}
\gamma^{(8)}_\a= \sigma_1\otimes \sigma_\a \otimes \mathbb{I}~,~~~
\gamma^{(8)}_a = \sigma_2\otimes \mathbb{I}\otimes \gamma_a~,~~~
B^{(8)}=\sigma_3\otimes \sigma_2\otimes B_5~,
\eeq
with the charge conjugation matrix satisfying $B_5*=-B_5$, $\a$ an index on $S^3$ and $a$ an index on $M_5$. The pseudoreal $Spin(5)$ spinor $\eta$ of unit norm gives rise to an $SU(2)$-structure on $M_5$, via \cite{Apruzzi:2015zna}
\beq\label{su25}
\eta\otimes \eta^{\dag}= \frac{1}{4}(1+ V)\wedge e^{-i J},~~~~\eta\otimes \eta^{c\dag}= \frac{1}{4}(1+V)\wedge \omega~.
\eeq
The (local) $SU(2)$-structure consists of a real one-form $V$, real two-form $J$ and a complex two-form $\omega$ such that $J \wedge J = \frac12 \o \wedge \o^*$ and $\iota_V J = \iota_V \omega = J \wedge \omega = \omega \wedge \omega = 0$. \\

By making use of this decomposition, the $Spin(7)$ four-form decomposes in terms of the $\mathbb{I} \times SU(2)$-structure as
\begin{align}\label{spin7decomp}
\Psi &=-\frac{1}{2}J \wedge J + \frac{e^{\tilde C}}{2} V \wedge (K_1\wedge \text{Re}\omega + K_2\wedge \text{Im}\omega + K_3\wedge J )\\[2mm]
&- \nu \frac{e^{2\tilde C}}{4} (dK_1\wedge \text{Re}\omega + dK_2\wedge \text{Im}\omega +dK_3\wedge J )- \nu e^{3\tilde C} V\wedge \text{Vol}(S^3) ~.
\end{align}
Using the decomposed $\Psi$, we examine the supersymmetry conditions. First, we consider the flux component $F$, which we decompose as
\eq{
F =  e^{3 C} \text{Vol}(S^3) \wedge F_1 + \star_5 \tilde{F}_1 ~,
}
with $F_1, \tilde{F}_1$ one-forms on $M_5$. Inserting this and \eqref{spin7decomp} into \eqref{27}, it follows from $(m,n) = (a,b)$ that $F_1 \sim \tilde{F}_1$, $F_1^a V_a = 0$ and from $(m,n) = (\a,a)$ that $F_1^a \omega_{ab} = F_1^a \omega_{ab}^* = F_1^a J_{ab} = 0$. Hence $F_1 = \tilde{F}_1 = 0$, hence $F = 0$. The equation of motion for the flux \eqref{tadpole} thus reduces to the following constraint on the warp factors:
\eq{
d(e^{3 C} \star_5 d(e^{-3\Delta})) = 0 ~.
}
Next, the requirement that $M_8$ is of $Spin(7)$ holonomy is equivalent to the closure of $\Psi$, which is equivalent to
\beq
d(e^{3\tilde C}V)= d(e^{2\tilde C}J)+2\nu e^{\tilde C}V\wedge J=d(e^{2\tilde C}\omega)+2\nu e^{\tilde C}V\wedge \omega= d(J\wedge J)=0~.
\eeq
In general, this means that locally $V = e^{- 3 C} d \tau$, and we will write
\eq{
ds^2(M_8) = e^{2 C} ds^2(S^3) + e^{-6 C} d \tau^2 + ds^2_4
}
Let us give some simple classes of examples for which the above conditions are solved.\\

$\bullet$ In the case that $C = C (\tau)$, the metric $ds_4^2$ is conformally Calabi-Yau. Let $J = e^{2 (W - C)} \tilde{J}$,  $\o = e^{2 (W - C)} \tilde{ \o}$. Then provided that we define $W (\tau) $ to satisfy
\eq{
W' + \nu e^{- 4 C} = 0 ~,
}
we find that $d \tilde{J} = d \tilde{\o} = 0$.
\\\\
$\bullet$ By taking $C = - \frac14 \log (4 \tau)$, we find that $d J = d \o = 0$, hence $ds_4^2$ is a Calabi-Yau metric\footnote{We have redefined $\tau$ to absorb the sign $\nu$.}. Introducing $\rho = (4 \tau)^{1/4}$, the metric reduces to
\eq{
ds^2(M_8) = d \rho^2 + \r^2 e^{2 C} ds^2(S^3)  + ds^2_4 ~,
}
and thus  $M_8 = \rbb^4 \times Y_2$, where $Y_2$ is a Calabi-Yau two-fold.\\

$\bullet$ Next, let us examine Sasaki-Einstein structures, as well as a class of generalizations. It can be shown that any five-dimensional Sasaki-Einstein can be defined by means of a set of real forms $(\tilde{V}, \o_j)$, $j=1,2,3$ with $\tilde{V}$ a one-form and $\o_j$ two-forms. These satisfy  \cite{lt3}
\eq{
d \o_1 = d (\o_2 + i \o_3 ) + 3 i \tilde{V} \wedge (\o_2 + i \o_3 ) = 0~.
}
A more general class of spaces are the so-called hypo manifolds \cite{cs}, satisfying $ d \o_1 = d (\tilde{V} \wedge \o_2 ) = d (\tilde{V} \wedge \o_3 ) = 0$, which themselves are a subclass of balanced manifolds \cite{mtuv}, satisfying
\eq{
d (\o_1 \wedge \o_1) = d (\tilde{V} \wedge \o_2 ) = d (\tilde{V} \wedge \o_3 ) = 0 ~.
}
By setting $J= \o_1$, $ e^c V = \tilde{V}$, $ \text{Re} \o = \o_2$, $\text{Im}\o = \o_3$, it follows that any solution to the supersymmetry constraints is a balanced metric. On the other hand, any solution to the supersymmetry constraints which is hypo automatically is such that $ds_4^2$ is Calabi-Yau. This leads to the conclusion that the spinors do not define a Sasaki-Einstein on $M_5$, as the base space of a Sasaki-Einstein manifold is not Ricci-flat. Another way to see this is to note that Sasaki-Einstein metrics can be written as a fibration over a K\"{a}hler-Einstein base, but it is clear that since the supersymmetry constraints are invariant under permutations of $(J, \text{Re}\o, \text{Im} \o)$, $ds_4^2$ cannot be non-Calabi-Yau K\"{a}hler.

\subsection{$G_2$-structure}
Next, let us examine the case where both internal chiral Killing spinors are (locally) non-vanishing. Again following \cite{Martelli:2003ki},
the Killing spinor is given by
\eq{
\e = e^{- \Delta} \theta \otimes \left( \chi_+ + \chi_- \right)~,
}
leading to the solution
\beq
ds^2= e^{2\Delta}\bigg(ds^2(\mathbb{R}^{1,2})+ds^2(M_8)\bigg)~,~~~  G=e^{3\Delta}( \text{Vol}_3\wedge f+ F).
\eeq
This time, the metric is not of special holonomy. Instead, it allows a $G_2$-structure with non-trivial torsion.
The norms of $\chi_{\pm}$ can be parametrised as
\beq
|\chi_{\pm}|^2=1\pm \sin\zeta~,
\eeq
with $\sin \z $ a function of $M_8$ such that the norms of $\chi_\pm$ are non-vanishing. The bilinears of $\chi_\pm$ defining the $G_2$-structure are given by
\beq\label{g2struc}
K_m = \frac{1}{\cos\zeta}\chi^{\dag}_+\gamma^{(8)}_m\chi_{-}, ~~~ \varphi_{mnp} = \frac{1}{\cos\zeta}\chi^{\dag}_+\gamma^{(8)}_{mnp}\chi_{-},
\eeq
In terms of these, the constraints
\begin{subequations}
\begin{align}
d(e^{3\Delta}\cos\zeta K)&=0~,\label{eq:Mtheorybpsa}\\[2mm]
K\wedge d(e^{6\Delta}\iota_K\star_8\varphi)&=0~,\\[2mm]
d(e^{12\Delta}\cos\zeta\,\varphi\wedge\iota_K\star_8\varphi)&=0~,\\[2mm]
\cos\zeta d(\varphi)\wedge \varphi +4\star_8 d\zeta-2 \cos\zeta \star_8 f&=0~,\\[2mm]
d(e^{3\Delta}\sin\zeta)-e^{3\Delta}f= d(e^{6\Delta}\cos\zeta\varphi)+e^{6\Delta}\star_8 F-e^{6\Delta}\sin\zeta F&=0\label{eq:Mtheorybpsb}.
\end{align}
\end{subequations}
are locally equivalent to the supersymmetry conditions.\\

Next, we split $M_8= S^3\times M_5$, leading to the following decomposition of the metric and flux:
\beq
ds^2(M_8)= e^{2C}ds^2(S^3) + ds^2(M_5)~,~~~~ F= e^{3C}F_1\wedge\text{Vol}(S^3)  + F_4~.
\eeq

The spinors decompose as
\beq
\chi_{+}=\sqrt{1+ \sin\zeta} \left(\begin{array}{c}1\\0\end{array}\right)\otimes( \xi\otimes \eta_{1}+ \xi^c\otimes \eta^c_{1})~,~~~
\chi_{-}=\sqrt{1- \sin\zeta} \left(\begin{array}{c}0\\1\end{array}\right)\otimes( \xi\otimes \eta_{2}+\xi^c\otimes \eta^c_{2})
\eeq
and the gamma matrices again decompose as \eqref{eq:gammasforMtheory}. The $Spin(5)$ spinors can be expanded in a common basis as
\beq
\eta^1= \eta,~~~\eta^2= a_0\eta+ a\eta^c+ \frac{b}{2}\overline{w}\eta,~~~|a_0|^2+|a|^2+|b|^2=1
\eeq
where $b$ can be made real by rotating the 1-form $w$ and $\eta$ is unit norm. We assume $w = w_1 + i w_2 $ is locally non-vanishing.\\
As a result, a second locally non-vanishing 1-form can be defined as
\eq{
u=\frac{1}{2}\iota_{w^*}\omega ~,
}
with $\omega$ defined as in \eqref{su25}.
We thus see that the local $SU(2)$-structure defined on $M_5$ by $\eta$ reduces further to a trivial structure, with the local vielbein defined by $(V, w_1, w_2, u_1, u_2)$.\\

We now express the $G_2$-structure \eqref{g2struc} in terms of the trivial structure of $S^3 \times M_5$. The first bilinear we calculate is
\beq
K = \frac{e^{C}}{2}(\text{Im}a_0K_1+\text{Re}a_0K_2- \text{Im}aK_3)-b u_2- \text{Re} a V,
\eeq
and the only way to make this compatible with \eqref{eq:Mtheorybpsa} is to set
\beq
a_0=\text{Im}a=0
\eeq
so that the spinors $\eta_{1,2}$ are nowhere parallel. We are now free to parametrise
\beq
b= \cos\alpha,~~~ \text{Re} a =\sin\alpha
\eeq
and rotate to a frame where
\beq
K= V.
\eeq
Having done this the other bilinears take the form
\begin{align}
\varphi&= -\cos\alpha e^1\wedge e^2\wedge e^3 - e^{3C}\sin\alpha\text{Vol}(S^3)-\nu\frac{e^{2C}}{4}\cos\alpha e^i\wedge dK_i,\nn\\[2mm]
&- \frac{e^{C}}{2}K_i\wedge ( u_1\wedge e^i+ \frac{1}{2}\epsilon_{ijk}\sin\a\,e^{j}\wedge e^{k}),\nn\\[2mm]
\iota_{K}\star_8\varphi&=u_1\wedge (\sin\alpha e^1\wedge e^2\wedge e^3 - e^{3C}\cos\alpha \wedge\text{Vol}(S^3))+\nu\frac{e^{2C}}{4}(\sin\a\,u_1\wedge e^i+\frac{1}{2}\epsilon_{ijk} e^j\wedge e^{k})\wedge dK_i\nn\\[2mm]
&- \frac{e^{C}}{4}\cos\alpha\epsilon_{ijk}u_1\wedge e^j\wedge e^k\wedge K_i .
\end{align}
where we have defined
\beq
e=(w_1,~w_2,~-u_2),
\eeq
for ease of presentation. Remark that
\begin{equation}
\varphi \wedge \iota_{K}\star_8\varphi = 7\nu\, \textrm{Vol}_7,
\end{equation}
where $\textrm{Vol}_7$ is the volume form of the manifold spanned by the warped left-invariant forms of $S^3$ and the vielbein, with orientation
\[\left\{\frac{e^C}{2}K_1,\frac{e^C}{2}K_2,\frac{e^C}{2}K_3,u_1, e^1,e^2,e^3\right\}.\]
Inserting these definitions for $\varphi$ and $\iota_{K}\star_8\varphi$ into \eqref{eq:Mtheorybpsa}- \eqref{eq:Mtheorybpsb} lead to the 5d conditions
\begin{subequations}\label{mtheorysusy}
	\begin{align}
	&d(e^{3\Delta}\cos\zeta V)=d(e^{6\Delta+3C}\cos\a u_1)\wedge V=d(\frac{e^{-6\Delta-3C}}{\cos^3\alpha \cos^2\zeta}u_1 )\wedge e^1\wedge e^2\wedge e^3=0~, \label{eq:MtheorySUSYa}\\[2mm]
	&d(e^{6\Delta+2C}\cos\zeta \cos\alpha e^i)+ \nu \,e^{6\Delta+C}\cos\zeta(2 u_1\wedge e^i+ \sin\alpha \epsilon_{ijk}e^j\wedge e^k)=0~, \label{eq:MtheorySUSYb}\\[2mm]
	& \!\!\!\bigg(d(e^{6\Delta+2C}(\sin\alpha u_1\wedge e^i\!+\!\frac{1}{2} \epsilon_{ijk}e^{j}\wedge e^k))\!+\!\nu e^{6\Delta+C}\cos\alpha \epsilon_{ijk}u_1\wedge e^{j}\wedge e^k\bigg)\wedge V= 0~, \label{eq:MtheorySUSYc}\\[2mm]
	&d(e^{-2C}\cos\alpha \epsilon_{ijk}  e^{j})\wedge e^k\wedge u_1\wedge V=\epsilon_{ijk} u_1\wedge du_1\wedge e^j\wedge e^k=0~, \label{eq:MtheorySUSYd}\\[2mm]
	&-2\cos\zeta d\alpha\wedge e^1\wedge e^2\wedge e^3+ 2 \star_5 d\zeta-\cos\zeta\star_5 f=d(e^{3\Delta}\sin\zeta)-e^{3\Delta}f=0~, \label{eq:MtheorySUSYe}\\[2mm]
	&d(e^{6\Delta+3C}\cos\zeta\sin\alpha)+e^{6\Delta+3C}( -\star_5 F_4+\sin\zeta F_1)=0~, \label{eq:MtheorySUSYf}\\[2mm]
	&d(e^{6\Delta}\cos\zeta \cos\alpha e^1\wedge e^2\wedge e^3)+ e^{6\Delta}(\star_5F_1  - \sin\zeta F_4)=0~, \label{eq:MtheorySUSYg}
	\end{align}
\end{subequations}
where we have used that $\cos\zeta\neq 0$ and one can show that $\cos\alpha= 0$ is inconsistent with supersymmetry.
In addition, the Bianchi identities and equations of motion for the flux reduce to
\eq{
d \left( e^{3 \Delta} f \right) = d \left( e^{3 \Delta+3 C} F_1 \right)= d \left( e^{3 \Delta} F_4 \right) &= 0 \\
d \left( e^{6 \Delta + 3 C} \star_5 F_4 \right) - e^{6 \Delta + 3 C} f \wedge F_1 &= 0 \\
d \left( e^{6 \Delta} \star_5 F_1 \right)       - e^{6 \Delta} f \wedge F_4 &= 0 \\
d \left( e^{6 \Delta+ 3 C} \star_5 f   \right)  + e^{6 \Delta + 3 C} F_1 \wedge F_4 &= 0 ~.
}
Note that the signs are such that supersymmetry together with the Bianchi identities imply the first two equations of motions. All Hodge duals in the above are with respect to the unwarped five-dimensional metric.

\end{document}

%% file: niall-macros.tex
\newcommand{\beq}{\begin{equation}}
\newcommand{\eeq}{\end{equation}}
\newcommand{\bea}{\begin{eqnarray}}
\newcommand{\eea}{\end{eqnarray}}
\newcommand{\nn}{\nonumber}

%% file: daniel-macros.tex
\def\a{\alpha}
\def\b{\beta}

\def\e{\epsilon}

\def\F{\Phi}
\def\g{\gamma}
\def\G{\Gamma}

\def\l{\lambda}

\def\m{\mu}
\def\n{\nu}
\def\o{\omega}
\def\O{\Omega}

\renewcommand{\t}{\theta}
\def\r{\rho}
\def\s{\sigma}

\def\z{\zeta}
\def\p{\partial}
\def\obb{\mathbb{I}}
\def\rbb{\mathbb{R}}
\newcommand{\ncal}{\mathcal{N}}

\newcommand{\eq}[1]{\begin{equation}\begin{split}#1\end{split}\end{equation}}
\renewcommand{\d}{\text{d}}

\newcommand{\arxth}[1]{\href{http://arxiv.org/abs/hep-th/#1}{[{\tt hep-th/#1}]}}
\newcommand{\arxdg}[1]{\href{http://arxiv.org/abs/math/#1}{\tt math.dg/#1}}
\newcommand{\arx}[1]{[\href{http://arxiv.org/abs/#1}{\tt #1}]}